%% file: TheATLASJulyPaper.tex
\documentclass[final,3p,times,twocolumn]{elsarticle}
\usepackage{graphicx}
\usepackage{bm}
\usepackage{atlasphysics} 
\usepackage{multirow}
\usepackage{amsmath}
\usepackage{float}
\usepackage{subfigure}
\usepackage{rotating}
\usepackage{wasysym}

\journal{Physics Letters B}

\usepackage{preprintcover}  
\PreprintCoverPaperTitle{\papertitle}  
\PreprintIdNumber{CERN-PH-EP-2012-218}  
\PreprintCoverAbstract{A 
  search for the Standard Model Higgs boson in
  proton-proton collisions with the ATLAS detector at the LHC is
  presented.  The datasets used correspond to integrated luminosities
  of approximately 4.8~\ifb\ collected at $\sqrt{s}=7$\,TeV in 2011 and 5.8~\ifb\
  at $\sqrt{s}=8$\,TeV in 2012.  Individual searches in the channels
  \htollll, \hgg\ and $\hWWenmun$ in the
  8\,TeV data are combined with previously published results of
  searches for $H{\rightarrow\,}ZZ^{(*)}$, $WW^{(*)}$, \bbbar\ and
  $\tau^+\tau^-$ in the 7\,TeV data and results from improved analyses of the \htollll\ and
  \hgg\  channels in the 7\,TeV data.  Clear evidence
  for the production of a neutral boson with a measured mass of $\massresultStatSys$ 
  is presented. This observation, which has a significance of 5.9 standard deviations,
  corresponding to a background fluctuation probability of $1.7\times 10^{-9}$, 
  is compatible with the production and decay of the Standard Model Higgs boson. 
}  
\PreprintJournalName{Physics Letters B}  

\newcommand{\metrel}{\ensuremath{E_{\rm T,rel}^{\rm miss}}}
\newcommand{\mT}{\ensuremath{m_{\rm T}}}
\newcommand{\pTll}{\ensuremath{p_{\rm T}^{\ell\ell}}}
\newcommand{\ptll}{\pTll}
\newcommand{\vpTll}{\ensuremath{{\bf p}_{\rm T}^{\ell\ell}}}

\newcommand{\lelm}{\ensuremath{H{\rightarrow\,}WW^{(\ast)}{\rightarrow\,}e \nu \mu \nu}}

\newcommand{\ZeroJet}{\mbox{0-jet}}
\newcommand{\OneJet}{\mbox{1-jet}}
\newcommand{\TwoJet}{\mbox{2-jet}}
\newcommand{\Wg}{\ensuremath{W\gamma^{(\ast)}}}

\newcommand{\hwwmll}{\ensuremath{m_{\ell\ell}}}

\def\WWSFzerojet{\ensuremath{1.06 \pm 0.06}} 
\def\WWSFonejet{\ensuremath{0.99 \pm 0.15}}

\def\TopSFonejet{\ensuremath{1.11 \pm  0.05}}
\def\TopSFtwojet{\ensuremath{1.01 \pm  0.26}}

\def\CaloMETCutem{\ensuremath{25}}

\def\IsoCutRange{0.12-0.20}
\def\IsoConeSize{\ensuremath{0.3}}

\def\WjetsSRErrorEle{\ensuremath{40\%}}

\newcommand{\atlasnote}[1]{\def\@atlasnote{#1}}

\newcommand{\ttH}{\ensuremath{q\bar{q}/gg \to t\bar{t}H}}

\newcommand{\Zjets}{$Z$+jets}

\newcommand{\ptt}{\ensuremath{p_{\mathrm{Tt}}}}
\newcommand{\hgg}{\ensuremath{H{\rightarrow\,}\gamma\gamma}}

\newcommand{\ggWW}{\ensuremath{gg{\rightarrow\,}WW}}
\newcommand{\hWWlnln}{\ensuremath{H{\rightarrow\,}WW^{(*)}{\rightarrow\,}\ell\nu\ell\nu}}

\newcommand{\hwwlnln}{\ensuremath{H{\rightarrow\,}WW^{(*)}{\rightarrow\,}\ell\nu\ell\nu}}

\newcommand{\hWWenmun}{\ensuremath{H{\rightarrow\,}WW^{(*)}{\rightarrow\,}e\nu\mu\nu}}

\newcommand{\hwwenmun}{\ensuremath{H{\rightarrow\,}WW^{(*)}{\rightarrow\,}e\nu\mu\nu}}

\newcommand{\hZZllll}{\ensuremath{H{\rightarrow\,}ZZ^{(*)}{\rightarrow\,}4\ell}}

\newcommand{\htt}{\ensuremath{H \rightarrow \tau^+\tau^-}}

\newcommand{\mh}{\ensuremath{m_{H}}}

\newcommand{\ztoee}{\ensuremath{Z{\rightarrow\,}e^+e^-}}

\newcommand\htollll{$H{\rightarrow\,}ZZ^{(*)}{\rightarrow\,}4\ell$}

\newcommand{\Wjets}{$W$+jets}

\newcommand{\ttoWb}{t{\rightarrow\,}Wb}

\newcommand{\infb}{fb$^{-1}$}

\renewcommand{\ttbar}{\ensuremath{t\overline{t}}}

\newcommand{\lowerExpNoGeV}{110}
\newcommand{\lowerExp}{\lowerExpNoGeV\,GeV}
\newcommand{\upperExpNoGeV}{582}
\newcommand{\upperExp}{\upperExpNoGeV\,GeV}

\newcommand{\lowerlowerObsNoGeV}{111}

\newcommand{\upperlowerObsNoGeV}{122} 
\newcommand{\upperlowerObs}{\upperlowerObsNoGeV\,GeV}

\newcommand{\lowerObsNoGeV}{131}
\newcommand{\lowerObs}{\lowerObsNoGeV\,GeV}
\newcommand{\upperObsNoGeV}{559}
\newcommand{\upperObs}{\upperObsNoGeV\,GeV}

\newcommand{\significance}{$6.0\,\sigma$}
\newcommand{\expectedsignificance}{$4.9\,\sigma$}
\newcommand{\significanceESS}{$5.9\,\sigma$}

\newcommand{\masspeak}{126.0}

\newcommand{\massresultStatSys}{\mbox{\ensuremath{\masspeak \; \pm 0.4 \;\textrm{(stat)} \; \pm 0.4\; \textrm{(sys)}~\mathrm{GeV}}}}

\usepackage{amssymb}

\usepackage{lineno}
\usepackage{chngpage}

\biboptions{numbers,square,comma,sort&compress}

\usepackage{hypernat}
\usepackage{hyperref}

\newcommand{\papertitle}{Observation of a New Particle in the Search for
  the Standard Model Higgs Boson with the ATLAS Detector at the LHC}

\begin{document}

\begin{frontmatter}

\title{\papertitle}

\author{The ATLAS Collaboration\\ \vspace{1cm} This paper is dedicated to the memory of our ATLAS 
colleagues who did not live to see the full impact and significance of their
contributions to the experiment. \vspace{-0.75cm}
}

\begin{abstract}

  A 
  search for the Standard Model Higgs boson in
  proton-proton collisions with the ATLAS detector at the LHC is
  presented.  The datasets used correspond to integrated luminosities
  of approximately 4.8~\ifb\ collected at $\sqrt{s}=7$\,TeV in 2011 and 5.8~\ifb\
  at $\sqrt{s}=8$\,TeV in 2012.  Individual searches in the channels
  \htollll, \hgg\ and $\hWWenmun$ in the
  8\,TeV data are combined with previously published results of
  searches for $H{\rightarrow\,}ZZ^{(*)}$, $WW^{(*)}$, \bbbar\ and
  $\tau^+\tau^-$ in the 7\,TeV data and results from improved analyses of the \htollll\ and
  \hgg\  channels in the 7\,TeV data.  Clear evidence
  for the production of a neutral boson with a measured mass of $\massresultStatSys$ 
  is presented. This observation, which has a significance of 5.9 standard deviations,
  corresponding to a background fluctuation probability of $1.7\times 10^{-9}$, 
  is compatible with the production and decay of the Standard Model Higgs boson. 

\end{abstract}

\end{frontmatter}


\hyphenation{ATLAS}

\section{Introduction}

The Standard Model (SM) of particle
physics~\cite{np_22_579,prl_19_1264,sm_salam,tHooft:1972fi} has been
tested by many experiments over the last four decades and has been
shown to successfully describe high energy particle
interactions. However, the mechanism that breaks electroweak symmetry
in the SM has not been verified experimentally.  This
mechanism~\cite{Englert:1964et,Higgs:1964ia,Higgs:1964pj,Guralnik:1964eu,Higgs:1966ev,Kibble:1967sv},
which gives mass to massive elementary particles, implies the
existence of a scalar particle, the SM Higgs boson.  The search for
the Higgs boson, the only elementary particle in the SM that has
not yet been observed, is one of the highlights of the Large Hadron
Collider~\cite{1748-0221-3-08-S08001} (LHC) physics programme.

Indirect limits on the SM Higgs boson mass of $m_H<158\GeV$ at $95\%$
confidence level (CL) have been set using global fits to precision
electroweak results~\cite{lepew:2010vi}.  Direct searches at
LEP~\cite{Barate:2003sz}, the Tevatron~\cite{Aaltonen:2012if,D0combo072012,TevatronCombo:2012} and
the LHC~\cite{paper2012prd,Chatrchyan:2012tx} have previously
excluded, at 95\%\ CL, a SM Higgs boson with mass below
600\,GeV, apart from some mass regions between 116\,GeV and 127\,GeV.

Both the ATLAS and CMS Collaborations reported excesses of events in
their 2011 datasets of proton-proton ($pp$) collisions at
centre-of-mass energy $\sqrt{s}=7$\,TeV at the LHC, which were
compatible with SM Higgs boson production and decay in the mass region
124--126\,GeV, with significances of 2.9 and 3.1 standard
deviations ($\sigma$), respectively~\cite{paper2012prd,Chatrchyan:2012tx}.
The CDF and {D\O} experiments at the Tevatron have also recently
reported a broad excess in the mass region 120--135\,GeV; using the 
existing LHC constraints, the observed local significances for
$m_H=125$\,\GeV\ are 2.7\,$\sigma$ for CDF~\cite{Aaltonen:2012if}, 
1.1\,$\sigma$ for D{\O}~\cite{D0combo072012} and 
2.8\,$\sigma$ for their combination~\cite{TevatronCombo:2012}. 

The previous ATLAS searches in 4.6--4.8\,\infb\ of data at
$\sqrt{s}=7$\,TeV are combined here with new searches for \htollll\footnote{The symbol $\ell$ stands for
electron or muon.},
\hgg\ and $\hwwenmun$\ in the 5.8--5.9\,\infb\ of $pp$ collision data taken
at $\sqrt{s}=8$\,TeV between April and June 2012.
 
The data were recorded with instantaneous luminosities
up to $6.8\times 10^{33}~\mathrm{cm}^{-2}\mathrm{s}^{-1}$; they
are therefore affected by multiple $pp$ collisions occurring in the
same or neighbouring bunch crossings (pile-up). In the 7\,TeV
data, the average number of interactions per bunch crossing was approximately
10; the average increased to approximately 20 in the 8\,TeV data.
The reconstruction, identification and isolation criteria used for electrons and photons in the 8\,TeV data   
are improved, making the \hZZllll\ and  \hgg\ searches more robust against the increased pile-up. These analyses were re-optimised with simulation and
frozen before looking at the 8\,TeV data.

In the \hWWlnln\ channel, the
increased pile-up deteriorates the event missing transverse momentum, $\MET$, resolution,
which results in significantly larger Drell-Yan
background in the same-flavour final states. Since the $e\mu$ channel
provides most of the sensitivity of the search, only this
final state is used in the analysis of the 8\,TeV data. 
The kinematic region in which a
SM Higgs boson with a mass between 110~\GeV\ and 140~\GeV\ is searched for
was kept blinded during the analysis optimisation, 
until satisfactory agreement was found between the observed and 
predicted numbers of events in control samples dominated by the principal 
backgrounds.

This Letter is organised as follows.
The ATLAS detector is briefly described in Section~\ref{sec:detector}. 
The simulation samples and the signal predictions are presented in
Section~\ref{sec:papersamples}.
The analyses of the \hZZllll, \hgg\
and \hWWenmun\ channels are described in Sections~\ref{sec:h4l}--\ref{sec:hww}, 
respectively.  The statistical procedure used to
analyse the results is summarised in Section~\ref{sec:statproc}. The systematic uncertainties
which are correlated between datasets and search channels are described in Section~\ref{sec:CombSyst}.
The results of the combination of all channels are reported in Section ~\ref{sec:Results},
while Section~\ref{sec:Conclusion} provides the conclusions.
 
\section{The ATLAS detector\label{sec:detector}}

The ATLAS detector~\cite{ATLASLOI,ATLASTP,atlas-det} is a multipurpose particle physics
apparatus with forward-backward symmetric cylindrical geometry. The
inner tracking detector (ID) consists of a silicon pixel detector, a
silicon microstrip detector (SCT), and a straw-tube transition radiation tracker
(TRT).  The ID is surrounded by a thin superconducting solenoid which
provides a 2~T magnetic field, and by high-granularity liquid-argon
(LAr) sampling electromagnetic calorimetry.  The electromagnetic
calorimeter is divided into a central barrel (pseudorapidity\footnote{ATLAS
  uses a right-handed coordinate system with its origin at the nominal
  interaction point (IP) in the centre of the detector, and the
  $z$-axis along the beam line.  The $x$-axis points from the IP to
  the centre of the LHC ring, and the $y$-axis points upwards.
  Cylindrical coordinates $(r,\phi)$ are used in the transverse plane,
  $\phi$ being the azimuthal angle around the beam line.  Observables
  labelled ``transverse'' are projected into the $x-y$ plane.
  The pseudorapidity is defined in terms of the polar angle $\theta$ as
  $\eta=-\ln\tan(\theta/2)$.} 
$|\eta|<1.475$) and 
end-cap regions on either end of the detector ($1.375 < |\eta| <2.5$ for the outer wheel and $2.5< |\eta| <3.2$ for the
inner wheel). In the region matched to the ID ($|\eta|<2.5$), it is 
radially segmented into three layers. The first layer has a fine segmentation 
in $\eta$ to facilitate $e$/$\gamma$ separation from $\pi^0$ and to improve 
the resolution of the shower position and direction measurements.
In the region $|\eta|<1.8$, the electromagnetic calorimeter is
preceded by a presampler detector to correct for upstream energy losses.
An iron-scintillator/tile calorimeter gives hadronic coverage in the 
central rapidity range ($|\eta| < 1.7$), while a LAr hadronic end-cap calorimeter 
provides coverage over $1.5<|\eta|<3.2$. 
The forward regions~($3.2< |\eta| <4.9$) are 
instrumented with LAr calorimeters for both electromagnetic and 
hadronic measurements.
The muon spectrometer (MS) surrounds the
calorimeters and consists of three large air-core superconducting
magnets providing a toroidal field, each with eight coils, a system of
precision tracking chambers, and fast detectors for triggering. The
combination of all these systems provides charged particle measurements
together with efficient and precise lepton and photon
measurements in the pseudorapidity range $|\eta| < 2.5$. Jets and
$\MET$ are reconstructed using energy deposits over the
full coverage of the calorimeters, $|\eta| < 4.9$.
				   
\section{Signal and background simulation samples \label{sec:papersamples}}

The SM Higgs boson production processes considered in this analysis are the
dominant gluon fusion ($gg\to H$, denoted ggF), vector-boson
fusion ($qq'\to qq'H$, denoted VBF) and Higgs-strahlung
($qq'\to WH, ZH$, denoted $WH$/$ZH$).  
The small contribution from the associated production
with a \ttbar\ pair (\ttH, denoted $t\bar{t}H$) is taken into
account only in the \hgg\ analysis.

For the ggF process, the signal cross section is computed at up to next-to-next-to-leading order
(NNLO)
in QCD~\cite{Georgi:1977gs,Djouadi:1991tka,Dawson:1990zj,Spira:1995rr,Harlander:2002wh,Anastasiou:2002yz,Ravindran:2003um}. 
Next-to-leading order (NLO) electroweak
(EW) corrections are applied~\cite{Aglietti:2004nj,Actis:2008ug}, as
well as QCD soft-gluon re-summations at up to next-to-next-to-leading logarithm
(NNLL)~\cite{Catani:2003zt}. These calculations, which are described in
Refs.~\cite{Anastasiou:2008tj,deFlorian:2012yg,Anastasiou:2012hx,Baglio:2010ae}, assume
factorisation between QCD and EW corrections.
The transverse momentum, \pt, spectrum of the Higgs boson in the
ggF process follows the {\tt HqT}~calculation~\cite{deFlorian:2011xf}, which includes QCD corrections at NLO
and QCD soft-gluon re-summations up to NNLL;
the effects of finite quark masses are also taken into account~\cite{Bagnaschi:2011tu}.

For the VBF process, full QCD and EW corrections up to NLO~\cite{Cahn:1983ip,Ciccolini:2007jr,Ciccolini:2007ec,Arnold:2008rz} and
approximate NNLO QCD corrections ~\cite{Bolzoni:2010xr} are used to
calculate the cross section.  Cross sections of the
associated $WH/ZH$ processes ($VH$) are calculated including QCD corrections
up to NNLO~\cite{Glashow:1978ab,Han:1991ia,Brein:2003wg} and EW
corrections up to NLO~\cite{Ciccolini:2003jy}.  The cross sections for the
$t\bar{t}H$ process
are estimated up to NLO
QCD~\cite{Kunszt:1984ri,Beenakker:2001rj,Beenakker:2002nc,Dawson:2002tg,Dawson:2003zu}.

The total cross sections for SM Higgs boson production at the LHC with $m_H=125$\,GeV are predicted to
be 17.5 \,pb for $\sqrt{s}=7$\,TeV and 22.3\,pb for $\sqrt{s}=8$\,TeV~\cite{LHCHiggsCrossSectionWorkingGroup:2011ti,Dittmaier:2012vm}.

The branching ratios of the SM Higgs boson
as a function of $m_{H}$, as well as their uncertainties, are calculated using the
HDECAY~\cite{Djouadi:1997yw} and PROPHECY4F~\cite{Bredenstein:2006rh,Bredenstein:2006ha} programs
and are taken from
Refs.~\cite{LHCHiggsCrossSectionWorkingGroup:2011ti,Dittmaier:2012vm}.
The interference in the \htollll\ final states with identical leptons is taken into account~\cite{Bredenstein:2006rh,Bredenstein:2006ha,Dittmaier:2012vm}.

 \begin{table}[!htbp]
   \centering
   \caption{Event generators used to model the signal and background
     processes. ``PYTHIA'' indicates that PYTHIA6 and PYTHIA8
     are used for simulations of $\sqrt{s}=7$\,TeV and $\sqrt{s}=8$\,TeV data, respectively.
     }
   \vspace{0.3cm}
   \scalebox{0.95}{
   \begin{tabular}{ll}
     \hline\hline
     Process & Generator \\
     \hline
     ggF, VBF & POWHEG~\cite{Alioli:2008tz,Nason:2009ai}+PYTHIA \\
     $WH$, $ZH$, $t\bar{t}H$ & PYTHIA  \\
     \hline
     $W$+jets, $Z/\gamma^{*}$+jets  & ALPGEN~\cite{alpgen}+HERWIG \\
     $\ttbar$, $tW$, $tb$           & MC@NLO~\cite{mcatnlo}+HERWIG \\
     $tqb$                          & AcerMC~\cite{Kersevan:2004yg}+PYTHIA \\
     $q\bar{q}\rightarrow WW$       & MC@NLO+HERWIG  \\
     $gg\rightarrow WW$             & gg2WW~\cite{gg2WW}+HERWIG \\
     $q\bar{q}\to ZZ$               & POWHEG~\cite{Melia:2011tj}+PYTHIA \\
     $gg \to ZZ$                    & gg2ZZ~\cite{gg2ZZ}+HERWIG \\
     $WZ$                           & MadGraph+PYTHIA, HERWIG \\
     $W\gamma$+jets                 & ALPGEN+HERWIG \\
     $W\gamma^{*}$~\cite{Gray:2011us}  & MadGraph+PYTHIA \\
     $q\bar{q}/gg\to\gamma\gamma$   & SHERPA \\
     \hline\hline \\
   \end{tabular}
   } \label{tab:cross}
 \end{table}

The event generators used to model signal and background
processes in samples of Monte Carlo (MC) simulated events are listed in
Table~\ref{tab:cross}.  The normalisations of the generated samples are
obtained from the state of the art calculations described above.
Several different programs are used to
generate the hard-scattering processes.  To generate parton
showers and their hadronisation, and to simulate the underlying event~\cite{MCTUNE2010,MCTUNEMC11,MCTUNEPY8},
PYTHIA6~\cite{pythia} (for 7\,TeV samples and 8\,TeV samples produced
with MadGraph~\cite{Alwall:2007st,Alwall:2011uj} or AcerMC) or
PYTHIA8~\cite{pythia8} (for other 8\,TeV samples) are used.
Alternatively, HERWIG~\cite{Corcella:2000bw} or SHERPA~\cite{Gleisberg:2008ta} are 
used to generate and hadronise parton showers, with the HERWIG underlying
event simulation performed using JIMMY~\cite{jimmy}.
When PYTHIA6 or HERWIG are used, TAUOLA~\cite{Jadach:1993hs} and PHOTOS~\cite{Golonka:2005pn} are employed
to describe tau lepton decays and additional photon radiation from charged leptons, respectively.

The following parton distribution function (PDF) sets are used: CT10 ~\cite{Lai:2010vv}
for the POWHEG, MC@NLO, gg2WW and gg2ZZ samples;
CTEQ6L1~\cite{cteq6} for the PYTHIA8, ALPGEN, AcerMC, MadGraph, HERWIG and SHERPA samples; and
MRSTMCal~\cite{mrst} for the PYTHIA6 samples.

Acceptances and efficiencies are obtained mostly from full simulations of
the ATLAS detector~\cite{atlassim} using {\sc Geant4}~\cite{GEANT4}. These simulations
include a realistic modelling of the pile-up conditions observed in the
data. Corrections obtained from measurements in data are
applied to account for small differences between data and simulation (e.g.\
large samples of $W$, $Z$ and $J/\psi$ decays are used to derive scale
factors for lepton 
reconstruction and identification efficiencies).

\renewcommand{\labelitemi}{$-$}
\newcommand\htollllp{$H\to ZZ^{(*)}\to 4\ell$}
\newcommand\htollllbrief{$H\to ZZ^{(*)}\to 4\ell$}
\newcommand\progname[1]{{\sc #1}}
\newcommand\pval{\ensuremath{p_0}}
\def\brocket#1{\left\langle #1 \right\rangle}
\def\Figref#1{Figure~\ref{#1}} 
\def\figref#1{Fig.~\ref{#1}}   
\def\secref#1{Section~\ref{#1}}
\newcommand{\tabscript}[3]{%
  \setlength{\fboxrule}{0pt}%
  \fbox{\ensuremath{#1^{#2}_{#3}}}%
}
\newcommand\candtwelvemu{22~}
\newcommand\candtwelveemu{19~}
\newcommand\candtwelvemue{14~}
\newcommand\candtwelvee{14~}
\newcommand\candtwelvemix{33~}
\newcommand\candtwelvetotal{79~}
\newcommand\bkgtwelveexp{$XX\pm X$~}
\newcommand\candmu{33~}
\newcommand\candmix{32~}
\newcommand\cande{21~}
\newcommand\candtotal{86~}
\newcommand\candmuexp{22.4~$\pm$~X.X~}
\newcommand\candmixexp{33.7~$\pm$~X.X~}
\newcommand\candeexp{15.3~$\pm$~X.X~}
\newcommand\bkgexp{71.2~$\pm$~X.X~}
\newcommand\lumia{\ensuremath{4.8~\ifb}}
\newcommand\lumiFourMuona{\ensuremath{4.8~\ifb}}
\newcommand\lumiFourElectrona{\ensuremath{4.9~\ifb}}
\newcommand\lumiTwoMuonTwoElectrona{\ensuremath{4.8~\ifb}}
\newcommand\lumib{\ensuremath{5.8~\ifb}}
\newcommand\lumiFourMuonb{\ensuremath{5.8~\ifb}}
\newcommand\lumiFourElectronb{\ensuremath{5.9~\ifb}}
\newcommand\lumiTwoMuonTwoElectronb{\ensuremath{5.8~\ifb}}
\newcommand\sensitivityExpected{0.XX}
\newcommand\sensitivityObserved{0.XX}
\newcommand\sensitivityMass{\ensuremath{200\:\gev}}

\newcommand\pvaluelowexp{10.6\%}
\newcommand\sigmalowexp{\ensuremath{1.3}}
\newcommand\pvaluelow{1.6\%}
\newcommand\sigmalow{\ensuremath{2.1}}
\newcommand\pvaluelowm{\ensuremath{125\gev}}

\newcommand\pvaluelowmOld{\ensuremath{242\gev}}
\newcommand\pvaluelowOld{0.5\%}
\newcommand\sigmalowOld{2.6}

\newcommand\pvaluehighmOld{\ensuremath{125\gev}}
\newcommand\pvaluehighOld{1.1\%}
\newcommand\sigmahighOld{2.3}

\newcommand\pvaluelowmNew{\ensuremath{125.5\gev}}
\newcommand\pvaluelowNew{0.4\%}
\newcommand\sigmalowNew{2.7}

\newcommand\pvaluehighmNew{\ensuremath{266\gev}}
\newcommand\pvaluehighNew{3.5\%}
\newcommand\sigmahighNew{1.8}

\newcommand\pvaluelowmComb{\ensuremath{125\gev}}
\newcommand\pvaluelowComb{0.018\%}
\newcommand\sigmalowComb{3.6}

\newcommand\pvaluehighmComb{\ensuremath{266\gev}}
\newcommand\pvaluehighComb{1.9\%}
\newcommand\sigmahighComb{2.1}

\newcommand\globalSigmaLowComb{2.5}
\newcommand\globalpvalue{0.65\%}

\newcommand\pvaluehighexp{0.14\%}
\newcommand\sigmahighexp{\ensuremath{3.0}}

\newcommand\pvaluehigh{1.3\%}
\newcommand\sigmahigh{\ensuremath{2.2}}
\newcommand\pvaluehighm{\ensuremath{244\gev}}
\newcommand\pvaluehightwoexp{7.1\%}

\newcommand\sigmahightwoexp{\ensuremath{1.5}}
\newcommand\pvaluehightwo{1.8\%}
\newcommand\sigmahightwo{\ensuremath{2.1}}
\newcommand\pvaluehighmtwo{\ensuremath{500\gev}}

\newcommand\excludedrangeaBrief{131--162}
\newcommand\excludedrangebBrief{170--460}
\newcommand\excludedrangeexpaBrief{124--164}
\newcommand\excludedrangeexpbBrief{176--500}
\newcommand\excludedranges{$\excludedrangeaBrief\,\gev$ and $\excludedrangebBrief\,\gev$}
\newcommand\excludedrangesexp{$\excludedrangeexpaBrief\,\gev$ and $\excludedrangeexpbBrief\,\gev$}

\newcommand\excludedrangesJan{$134\!-\!156\,\gev$, $182\!-\!233\,\gev$, $256\!-\!265\,\gev$ and $268\!-\!415\,\gev$}
\newcommand\excludedrangesexpJan{$136\!-\!157\,\gev$ and $184\!-\!400\,\gev$}
\newcommand\lumiAverageJan{\ensuremath{4.8~\ifb}}

\section{\texorpdfstring{\htollllp{} channel}{H->ZZ(*)->4l channel}}
\label{sec:h4l}

  The search for the SM Higgs boson through the decay \htollllp{},
  where $\ell=e\text{ or }\mu$, provides good sensitivity
  over a wide mass range ($110$-$600\,\gev$),~largely due to the excellent momentum resolution
  of the ATLAS detector. This analysis searches for Higgs boson
  candidates by selecting two pairs of isolated leptons, each of which is comprised of
  two leptons with the same flavour and opposite charge.
  The expected cross section times branching ratio for the process \htollllp{} with
  $\mH=125\,\gev$ is 2.2\,fb for $\sqrt{s}=7\,\tev$ and 2.8\,fb for $\sqrt{s}=8\,\tev$.

  The largest background comes from continuum
  $(Z^{(*)}/\gamma^{*}) (Z^{(*)}/\gamma^{*})$ production, referred to
  hereafter as $ZZ^{(*)}$. For low masses there are also important
  background contributions from $Z+\rm{jets}$ and $t\bar{t}$
  production, where charged lepton candidates arise either from decays
  of hadrons with $b$- or $c$-quark content or from mis-identification
  of jets.

  The $7\,\tev$ data have been re-analysed and combined with the
  $8\,\tev$ data. The analysis is improved in several aspects
  with respect to~Ref.~\cite{ATLAS:2012ac} to enhance the sensitivity
  to a low-mass Higgs boson. In particular, the kinematic selections
  are revised, and the $8\,\tev$ data analysis benefits from
  improvements in the electron reconstruction and
  identification. The expected signal significances for a Higgs boson with
  $\mH=125\,\gev$ are 
  1.6~$\sigma$ for the $7\,\tev$ data (to be compared
  with 1.25~$\sigma$ in~Ref.~\cite{ATLAS:2012ac}) and 2.1~$\sigma$ for the 
  $8\,\tev$ data.
  
  \subsection{Event selection} 

  The data are selected using single-lepton
  or dilepton triggers. For the single-muon trigger, the \pt~threshold is $18\,\gev$ for the $7\,\tev$ data
  and $24\,\gev$ for the $8\,\tev$ data, while for the
  single-electron trigger the transverse energy, \et,~threshold 
  varies from $20\,\gev$ to $22\,\gev$ for the $7\,\tev$ data 
  and is $24\,\gev$ for the $8\,\tev$ data. For the dielectron triggers, the
  thresholds are $12\,\gev$ for both electrons. For
  the dimuon triggers, the thresholds for the $7\,\tev$ data are
  $10\,\gev$ for each muon, while for the $8\,\tev$ data the thresholds are 
  $13\,\gev$. An additional asymmetric dimuon trigger is used in the $8\,\tev$ data 
  with thresholds  $18\,\gev$ and $8\,\gev$ for the leading and sub-leading muon, respectively.

  Muon candidates are formed by matching reconstructed ID
  tracks with either a complete track or a track-segment reconstructed in the
  MS~\cite{ATLAS-CONF-2011-063}. The muon
  acceptance is extended with respect to Ref.~\cite{ATLAS:2012ac} 
  using tracks reconstructed in the forward region
  of the MS  ($2.5<|\eta|<2.7$), which is outside the ID coverage. 
  If both an ID and a complete MS track are 
  present, the two independent momentum measurements are combined;
  otherwise the information of the ID or the MS is used alone.
  Electron candidates must have a
  well-reconstructed ID track pointing to an electromagnetic calorimeter cluster
  and the cluster should satisfy a set of identification
  criteria~\cite{Aad:2011mk} that require the longitudinal and
  transverse shower profiles to be consistent with those expected for
  electromagnetic showers. Tracks associated with
  electromagnetic clusters are fitted using a Gaussian-Sum
  Filter~\cite{GSFConf}, which allows for
  bremsstrahlung energy losses to be taken into account.

  Each electron (muon) must satisfy $\pt>7\,\gev$ ($\pt>6\,\gev$) and
  be measured in the pseudorapidity
  range~$|\eta|<2.47$~($|\eta|<2.7$). 
  All possible quadruplet combinations with same-flavour opposite-charge lepton pairs are then formed.
  The most energetic lepton in the
  quadruplet must satisfy $\pt>20\,\gev$, and the second (third)
  lepton in $\pt$ order must satisfy $\pt >15\,\gev$ ($\pt
  >10\,\gev$). At least one of the leptons must satisfy the single-lepton 
  trigger or one pair must satisfy the dilepton trigger
  requirements. The leptons are required to be separated from each
  other by $\Delta R = \sqrt{(\Delta \eta)^{2} + (\Delta \phi)^{2}}>0.1$
  if they are of the same flavour and by $\Delta R>0.2$ otherwise. The
  longitudinal impact parameters of the leptons along the beam axis
  are required to be within $10\,{\rm mm}$ of the reconstructed
  primary vertex. The primary vertex used for the event is defined as the reconstructed
  vertex with the highest $\sum\pt^2$ of associated tracks and is
  required to have at least three tracks with $\pt > 0.4\GeV$. To
  reject cosmic rays, muon tracks are required to have a transverse
  impact parameter, defined as the distance of closest approach to
  the primary vertex in the transverse plane, of less than $1\,{\rm mm}$.

  The same-flavour and opposite-charge lepton pair
  with an invariant mass closest to the $Z$ boson mass ($m_{\Zboson}$)
  in the quadruplet is referred to as the leading lepton pair. Its invariant mass,
  denoted by $m_{12}$, is required to be between $50\,\gev$~and
  $106\,\gev$. The remaining same-flavour, opposite-charge lepton pair
  is the sub-leading lepton pair. Its invariant mass, $m_{34}$, is
  required to be in the range $m_{\rm{min}}<m_{34}<115\,\gev$, where
  the value of $m_{\rm{min}}$ depends on the reconstructed four-lepton
  invariant mass, $m_{4\ell}$. The value of $m_{\rm{min}}$ varies monotonically
  from $17.5\,\gev$ at $m_{4\ell}=120\,\gev$ to $50\,\gev$ at
  $m_{4\ell}=190\,\gev$~\cite{ATLAS-CONF-2012-092} and is constant above this value.  
  All possible lepton pairs in the quadruplet that have the 
  same flavour and opposite charge must satisfy
  $m_{\ell\ell}>5\,\gev$ in order to reject backgrounds involving the production and 
  decay of $J/\psi$ mesons. 
  If two or more quadruplets satisfy the above selection, the one with the highest
  value of $m_{34}$ is selected. Four different analysis sub-channels, $4e$,
  $2e2\mu$, $2\mu2e$ and $4\mu$, arranged by the flavour of the leading
  lepton pair, are defined.

  Non-prompt leptons from heavy flavour decays,
  electrons from photon conversions and jets mis-identified as electrons  
  have broader transverse impact parameter distributions than prompt leptons from 
  $\Zboson$ boson decays and/or are non-isolated. Thus, the 
  \Zjets\ and $t\bar{t}$~background contributions are
  reduced by applying a cut on the transverse impact parameter
  significance, defined as the transverse impact parameter divided by its
  uncertainty, $d_0/\sigma_{d_0}$. This is required to be less than 3.5 (6.5)
  for muons (electrons). The electron impact parameter is affected by
  bremsstrahlung and thus has a broader distribution. 
  
  In addition, leptons must satisfy
  isolation requirements based on tracking and calorimetric information. 
  The normalised track isolation discriminant
  is defined as the sum of the transverse momenta of tracks inside a
  cone of size $\Delta R=0.2$ around the lepton direction, excluding the lepton
  track, divided by the lepton $\pt$. The tracks considered in the sum
  are those compatible with the lepton vertex and have $\pt>0.4\,\gev$ ($\pt>1\,\gev$) 
  in the case of electron (muon) candidates. 
  Each lepton is required to have a normalised track isolation smaller than 0.15. 
  The normalised calorimetric isolation for electrons is
  computed as the sum of the $\et$ of positive-energy topological
  clusters~\cite{Lampl:1099735} with a reconstructed barycentre falling within a cone of
  size $\Delta R=0.2$ around the candidate electron cluster, divided by the
  electron $\et$. The algorithm for topological clustering suppresses noise by keeping cells
  with a significant energy deposit and their neighbours.
  The summed energy of the cells assigned to the electron cluster
  is excluded, while a correction is applied to account for the electron energy deposited outside the cluster. 
  The ambient energy deposition in the event from pile-up and 
  the underlying event is accounted for using a calculation of the median
  transverse energy density from
  low-\pt~jets~\cite{Cacciari:2007fd,Cacciari:2011ma}. The normalised
  calorimetric isolation for electrons is required to be less than 0.20. 
  The normalised calorimetric isolation discriminant for muons is defined
  by the ratio to the $\pt$ of the muon of the \et~sum of the calorimeter cells inside a
  cone of size  $\Delta R=0.2$ around the muon direction minus the energy deposited by the muon.
  Muons are required to have a normalised
  calorimetric isolation less than 0.30 (0.15 for muons without
  an associated ID track). For both the track- and calorimeter-based
  isolation, any contributions arising from other leptons of the
  quadruplet are subtracted.

  The combined signal reconstruction and selection efficiencies for a SM Higgs with $\mH=125\,\gev$
  for the $7\,\tev$ ($8\,\tev$) data are $37$\% ($36$\%) for the $4\mu$ channel,
  $20$\% ($22$\%) for the $2e2\mu$/$2\mu2e$ channels and $15$\%
  ($20$\%) for the $4e$ channel. 

  The $4\ell$ invariant mass resolution is improved by applying a
  $Z$-mass constrained kinematic fit to the leading lepton pair for
  $m_{4\ell}<190\,\gev$ and to both lepton pairs for higher masses. The
  expected width of the reconstructed mass distribution is
  dominated by the experimental resolution for $m_{H}<350\,\gev$,
  and by the natural width of the Higgs boson for higher masses ($30\,\gev$ at $m_{H} = 400\,\gev$). 
  The typical mass resolutions for $m_{H}=125\,\gev$ are $1.7\,\gev$, $1.7\,\gev$/$2.2\,\gev$ and
  $2.3\,\gev$ for the $4\mu$, $2e2\mu$/$2\mu2e$ and $4e$ sub-channels, respectively.

  \subsection{Background estimation}

  The expected background yield and composition are estimated
  using the MC simulation normalised to the theoretical cross section
  for $ZZ^{(*)}$ production and by methods using control regions from data for the
  $Z+\rm{jets}$ and $t\bar{t}$ processes. Since the background composition
  depends on the flavour of the sub-leading lepton pair, different
  approaches are taken for the $\ell\ell+\mu\mu$ and the $\ell\ell+ee$
  final states. The transfer factors needed to extrapolate the
  background yields from the control regions defined below to the signal
  region are obtained from the MC simulation. The MC description of the selection
  efficiencies for the different background components has been
  verified with data.

  The reducible $\ell\ell+\mu\mu$ background is dominated by
  $t\bar{t}$ and $Z+{\rm jets}$ (mostly $Zb\bar{b}$) 
  events. A control region is defined by removing the isolation
  requirement on the leptons in the sub-leading pair, and by requiring
  that at least one of the sub-leading muons fails the transverse impact
  parameter significance selection. These modifications remove
  $ZZ^{(*)}$ contributions, and allow both the $t\bar{t}$ and $Z+{\rm
  jets}$ backgrounds to be estimated simultaneously using a fit to the
  $m_{12}$ distribution. The $t\bar{t}$ background contribution is
  cross-checked by selecting a control sample of events 
  with an opposite charge $e\mu$ pair with an invariant mass
  between $50\,\gev$~and $106\,\gev$, accompanied by an opposite-charge
  muon pair. Events with a $\Zboson$ candidate decaying to a pair of
  electrons or muons in the aforementioned mass range are
  excluded. Isolation and transverse impact parameter significance requirements are applied
  only to the leptons of the $e\mu$ pair.
   
  In order to estimate the reducible $\ell\ell+ee$ background, a control region
  is formed by relaxing the selection criteria for the electrons of
  the sub-leading pair. The different sources of electron background
  are then separated into categories consisting of 
  non-prompt leptons from heavy flavour decays, electrons from photon
  conversions and jets mis-identified as electrons, using appropriate
  discriminating variables~\cite{Aad:2011rr}. 
  This method allows the sum of the $Z+{\rm jets}$ and $t\bar{t}$ background
  contributions to be estimated.
  As a cross-check, the same method is also
  applied to a similar control region containing same-charge sub-leading
  electron pairs. An additional cross-check of the $\ell\ell +ee$
  background estimation is performed by using a control region with
  same-charge sub-leading electron pairs, where the three highest $\pt$
  leptons satisfy all the analysis criteria whereas the selection cuts are relaxed
  for the remaining electrons. All the cross-checks yield
  consistent results. 

  \begin{table}[!htb]
  \centering
  \caption{Summary of the estimated numbers of $Z+{\rm jets}$ and $t\bar{t}$ background events, 
     for the $\sqrt{s}=7\,\tev$ and $\sqrt{s}=8\,\tev$ data in the entire phase-space of the analysis 
     after the kinematic selections described in the text.
   The backgrounds are combined for the $2\mu2e$ and $4e$ channels, as discussed in the text.
    The first uncertainty is statistical, while the second is systematic.\label{tab:bkg_overview}}
  \vspace{0.2cm}
  \scalebox{0.9}{
  \begin{tabular}{cr@{$\pm$}c@{$\pm$}lr@{$\pm$}c@{$\pm$}l}
    \hline\hline
    Background& \multicolumn{6}{c}{Estimated}  \\
    & \multicolumn{6}{c}{numbers of events}  \\
    & \multicolumn{3}{c}{$\sqrt{s}=7\,\tev$}&\multicolumn{3}{c}{$\sqrt{s}=8\,\tev$ }  \\
    \hline
    &\multicolumn{6}{c}{$4\mu$}\\ 
    \hline
     \Zjets               & 0.3&  0.1&  0.1& 0.5  &  0.1  &  0.2\\  
     $t\bar{t}$           &0.02&  0.02& 0.01& 0.04 &  0.02 &  0.02\\
    \hline
    &\multicolumn{6}{c}{$2e2\mu$}\\ 
    \hline
    \Zjets              &0.2  &  0.1   &  0.1  & 0.4  &   0.1  &  0.1\\
    $t\bar{t}$          &0.02  &  0.01  &  0.01 & 0.04 &  0.01  &  0.01\\
    \hline
    &\multicolumn{6}{c}{$2\mu2e$}\\ 
    \hline
    \Zjets, $t\bar{t}$     &2.6 & 0.4 & 0.4 & 4.9  &  0.8  &  0.7 \\
    \hline
    &\multicolumn{6}{c}{$4e$}\\ 
    \hline
    \Zjets, $t\bar{t}$     &3.1 & 0.6 & 0.5  & 3.9  &  0.7  &  0.8\\
    \hline
    \hline
  \end{tabular}}
\end{table}
  
\begin{figure}[!htb]
  \centering 
  \includegraphics[width=0.45\textwidth]{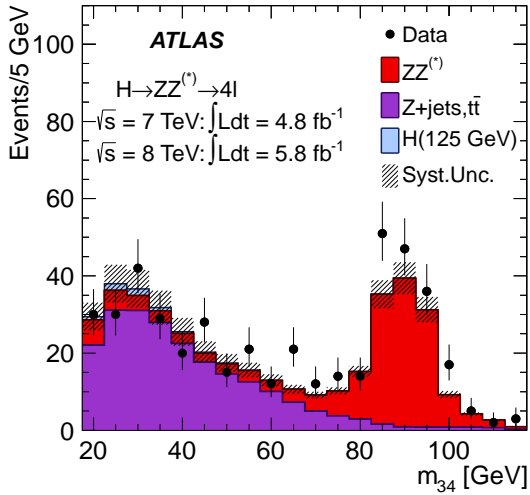}
  \caption{Invariant mass distribution of the sub-leading lepton pair ($m_{34}$) for a
  sample defined by the presence of a $Z$ boson candidate and an additional
  same-flavour electron or muon pair, for the combination of $\sqrt{s}=7\,\tev$ and~$\sqrt{s}=8\,\tev$~data
  in the entire phase-space of the analysis after the kinematic selections described in the text. 
  Isolation and transverse impact parameter significance requirements
  are applied to the leading lepton pair only. The MC is normalised to
  the data-driven background estimations. The relatively small contribution 
  of a SM Higgs with $m_{H}=125\,\gev$ in this sample is also shown.\label{fig:sub_cr_all}}
\end{figure}

  The data-driven background estimates are summarised
  in Table~\ref{tab:bkg_overview}. The distribution of $m_{34}$, for
  events selected by the analysis except that the isolation and
  transverse impact parameter requirements for the sub-leading lepton pair are removed, is
  presented in Fig.~\ref{fig:sub_cr_all}. 

  \subsection{Systematic uncertainties}\label{sec:h4lsyst}
  
  The uncertainties on the integrated luminosities are determined to be 1.8\%\ for the
  $7\,\tev$ data and 3.6\%\ for the $8\,\tev$ data using the techniques described in Ref.~\cite{ATLAS-CONF-2012-080}.

  The uncertainties on the lepton reconstruction and identification
  efficiencies and on the momentum scale and resolution are determined
  using samples of $W$, $Z$ and $J/\psi$ decays~\cite{Aad:2011mk,ATLAS-CONF-2011-063}. 
  The relative uncertainty on the signal acceptance due to the uncertainty on the
  muon reconstruction and identification efficiency is 
  $\pm 0.7\%$ ($\pm 0.5\%$/$\pm 0.5\%$) for
  the $4\mu$ ($2e2\mu$/$2\mu2e$) channel for $m_{4\ell}=600\,\gev$ and
  increases to $\pm 0.9\%$ ($\pm 0.8\%$/$\pm 0.5\%$) for
  $m_{4\ell}=115\,\gev$. Similarly, the relative uncertainty on the signal acceptance  
  due to the uncertainty on the electron reconstruction and identification efficiency 
  is $\pm 2.6\%$ ($\pm 1.7\%$/$\pm 1.8\%$) for
  the $4e$ ($2e2\mu$/$2\mu2e$) channel for $m_{4\ell}=600\,\gev$ 
  and reaches $\pm 8.0\%$ ($\pm 2.3\%$/$\pm 7.6\%$) for
  $m_{4\ell}=115\,\gev$. 
  The uncertainty on the electron energy scale
  results in an uncertainty of $\pm 0.7\%$ ($\pm 0.5\%$/$\pm
  0.2\%$) on the mass scale of the $m_{4\ell}$ distribution for the
  $4e$ ($2e2\mu$/$2\mu2e$) channel. 
  The impact of the uncertainties on the electron energy 
  resolution and on the muon momentum resolution and scale are found to be negligible.

  The theoretical uncertainties associated with the signal are described in detail 
  in Section~\ref{sec:CombSyst}.  
  For the SM $ZZ^{(*)}$ background, which is estimated from MC simulation, the uncertainty
  on the total yield due to the QCD scale uncertainty is
  $\pm 5\%$, while the effect of the PDF and $\alpha_{s}$
  uncertainties is $\pm 4\%$ ($\pm 8\%$) for processes 
  initiated by quarks (gluons)~\cite{Dittmaier:2012vm}. In addition,
  the dependence of these uncertainties on the four-lepton invariant mass spectrum has
  been taken into account as discussed in Ref.~\cite{Dittmaier:2012vm}.
  Though a small excess of events is observed for $m_{4l}>160\,\GeV$, the measured $ZZ^{(*)} \to 4\ell$  
  cross section~\cite{ATLAS-CONF-2012-090} is consistent 
  with the SM theoretical prediction.
  The impact of not using the theoretical constraints on the $ZZ^{(*)}$ yield 
  on the search for a Higgs boson with $\mH<2 m_{\Zboson}$
  has been studied in Ref.~\cite{ATLAS-CONF-2012-092} and
  has been found to be negligible .
  The impact of the interference between a Higgs signal and the non-resonant $gg \to ZZ^{(*)}$ 
  background is small and becomes negligible for $\mH<2 m_{\Zboson}$~\cite{Kauer:2012hd}.

  \begin{figure}[h!tp]
  \centering
  \includegraphics[width=0.45\textwidth]{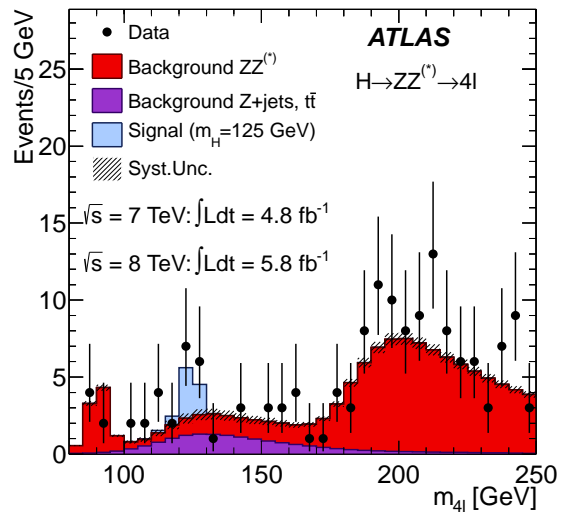}
  \caption{The distribution of the four-lepton invariant mass,
  $m_{4\ell}$, for the selected candidates, compared to the background
  expectation in the  $80$--$250\,\gev$
  mass range, for the combination of the $\sqrt{s}=7\,\tev$ and
  $\sqrt{s}=8\,\tev$ data. 
  The signal expectation for a SM Higgs with $\mH=125\,\gev$
  is also shown.\label{fig:finalMassesSignal}}
\end{figure} 

\subsection{Results}

\begin{figure}[h!tb]
  \centering
  \includegraphics[width=0.45\textwidth]{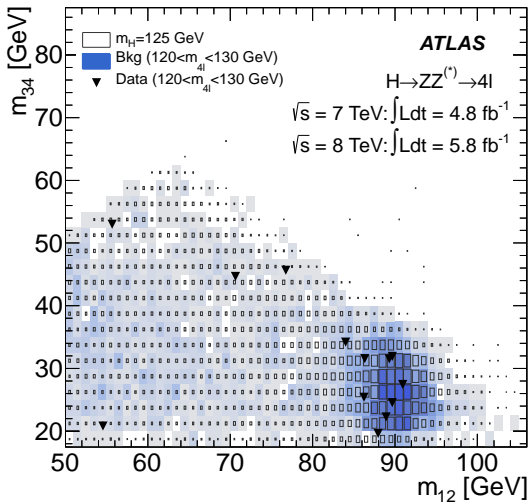}
 \caption{Distribution of the $m_{34}$ versus the $m_{12}$ invariant
    mass, before the application of the $Z$-mass constrained kinematic
    fit, for the selected candidates in the $m_{4\ell}$ range $120$--$130\,\gev$. 
    The expected distributions for a SM Higgs with $\mH=125\,\gev$ (the sizes of the boxes indicate the relative density) 
    and for the total background (the intensity of the shading indicates the relative density) are also shown.\label{fig:M12vsM34}}
\end{figure}

The expected distributions of $m_{4\ell}$ for the background and for a Higgs boson
 signal with $\mH=125\,\gev$ are compared to the data in Fig.~\ref{fig:finalMassesSignal}.
 The numbers of observed and expected events
 in a window of $\pm 5\,\gev$ around $\mH=125\,\gev$ are presented for
 the combined  $7\,\tev$ and $8\,\tev$ data in Table~\ref{tab:yields}. The distribution of the
 $m_{34}$ versus $m_{12}$ invariant mass is shown in Fig.~\ref{fig:M12vsM34}. The 
 statistical interpretation of
 the excess of events near $m_{4\ell}=125$\,GeV in Fig.~\ref{fig:finalMassesSignal} is presented in
 Section~\ref{sec:Results}.
  
\label{sec:h4lresults}
  \begin{table}[h!tb]
   \centering
    \caption{The numbers of expected signal ($\mH=125\,\gev$) and background events, together with the numbers of observed events in the data,
        in a window of size $\pm 5\,\gev$ around $125\,\gev$, for the combined $\sqrt{s}=7\,\tev$ and $\sqrt{s}=8\,\tev$ data.\label{tab:yields}}
    \vspace{0.3cm}
   \scalebox{0.85}{
    \begin{tabular}{@{}ccccc@{}}
    \hline\hline
    & Signal & $ZZ^{(*)}$ & $Z+\rm{jets}$,~$t\bar{t}$ & Observed \\
    \hline
    $4\mu$ & 2.09$\pm$0.30 & 1.12$\pm$0.05 & 0.13$\pm$0.04 & 6\\
    \hline
    $2e2\mu$/$2\mu2e$ &2.29$\pm$ 0.33 & 0.80$\pm$0.05 & 1.27$\pm$0.19 & 5\\
    \hline
    $4e$ & 0.90$\pm$0.14 & 0.44$\pm$0.04 & 1.09$\pm$0.20 & 2\\
    \hline\hline
    \end{tabular}}
  \end{table}	
 
\section{\texorpdfstring{\hgg\ channel}{H->gg channel}}
\label{sec:hgg}
The search for the SM Higgs boson through the decay \hgg\ is performed in the mass range between 110~GeV and 150~GeV.
The dominant background is SM diphoton production~($\gamma\gamma$);
contributions also come from $\gamma$+jet and jet+jet production
with one or two jets mis-identified as photons~($\gamma j$ and $jj$) and from the Drell-Yan process.
The $7\TeV$ data have been re-analysed and the results combined with those from the $8\TeV$ data.
Among other changes to the analysis, a new category of events with two jets is introduced, which enhances the sensitivity to the VBF process.
Higgs boson events produced by the VBF process have two forward jets, originating from the two scattered quarks, and tend
to be devoid of jets in the central region.
Overall, the sensitivity of the analysis has been improved by about $20\%$ with respect to that described in Ref.~\cite{ATLAS:2012ad}.

\subsection{Event selection}
\label{sec:hggselect}
The data used in this channel are selected using a diphoton trigger~\cite{Aad:2012xs},
which requires two clusters formed from energy depositions in the electromagnetic calorimeter.
An $E_{\mathrm{T}}$ threshold of $20\GeV$ is applied to each cluster for the $7\TeV$ data,
while for the $8\TeV$ data
the thresholds are increased to $35\GeV$ on the leading (the highest $E_{\mathrm{T}}$) cluster
and to $25\GeV$ on the sub-leading (the next-highest $E_{\mathrm{T}}$) cluster.
In addition, loose criteria are applied to the shapes of the clusters
to match the expectations for electromagnetic showers initiated by photons.
The efficiency of the trigger is greater than $99\%$ for events passing the final event selection.

Events are required to contain at least one reconstructed vertex with at least two associated tracks with $\pt>0.4$~GeV,
as well as two photon candidates.
Photon candidates are reconstructed in the fiducial region $|\eta|<2.37$, excluding the calorimeter barrel/end-cap transition region $1.37\leq|\eta|<1.52$.
Photons that convert to electron-positron pairs in the ID material
can have one or two reconstructed tracks matched to the clusters in the calorimeter.
The photon reconstruction efficiency is about $97\%$ for $E_{\mathrm{T}}>30$~GeV.

In order to account for energy losses upstream of the calorimeter and energy leakage outside of the cluster,
MC simulation results are used to calibrate the energies of the photon candidates; there are separate
calibrations for unconverted and converted candidates.
The calibration is refined by applying $\eta$-dependent correction factors,
which are of the order of $\pm 1\%$, determined from measured \ztoee\ events. 
The leading (sub-leading) photon candidate is required to have $E_{\mathrm{T}}>40$~GeV (30~GeV).

Photon candidates are required to pass identification criteria based on shower shapes in the electromagnetic calorimeter
and on energy leakage into the hadronic calorimeter~\cite{Aad:2010sp}.
For the $7\TeV$ data, this information is combined in a neural network,
tuned to achieve a similar jet rejection as the cut-based selection described in Ref.~\cite{ATLAS:2012ad}, but with higher photon efficiency.
For the $8\TeV$ data, cut-based criteria are used to ensure reliable photon performance 
for recently-recorded data.
This cut-based selection has been tuned to be robust against pile-up by relaxing requirements
on shower shape criteria more susceptible to pile-up, and tightening others.
The photon identification efficiencies, averaged over $\eta$, range from $85\%$ to above $95\%$ for the $E_{\mathrm{T}}$ range under consideration.

To further suppress the jet background, an isolation requirement is applied.
The isolation transverse energy is defined as the sum of the transverse energy of positive-energy
topological clusters, as described in Section~\ref{sec:h4l}, within a cone of size $\Delta R = 0.4$ around the photon candidate,
excluding the region within $0.125\times 0.175$ in $\Delta\eta\times\Delta\phi$ around the photon barycentre.
The distributions of the isolation transverse energy in data and simulation have been found to be in good agreement using electrons from 
\ztoee\ events 
and photons from $Z\to\ell^+\ell^-\gamma$ events.
Remaining small differences are taken into account as a systematic uncertainty.
Photon candidates are required to have an isolation transverse energy of less than $4\GeV$.

\subsection{Invariant mass reconstruction}
The invariant mass of the two photons is evaluated using the photon energies measured in the calorimeter,
the azimuthal angle $\phi$ between the photons as determined from the positions of the photons in the calorimeter, and the values of
$\eta$ calculated from the position of the identified primary vertex and the impact points of the photons in the calorimeter.

The primary vertex of the hard interaction is
identified by combining the following information in a global likelihood:
the directions of flight of the photons as determined
using the longitudinal segmentation of the electromagnetic calorimeter~(calorimeter pointing), the parameters of the beam spot,
and the $\sum\pt^2$ of the tracks associated with each reconstructed vertex.
In addition, for the $7\TeV$ data analysis, the reconstructed conversion vertex is used
in the likelihood for converted photons with tracks containing hits in the silicon layers of the ID.
The calorimeter pointing is sufficient to ensure that 
the contribution of the opening angle between the photons to the mass resolution is negligible.
Using the calorimeter pointing alone,
the resolution of the vertex $z$ coordinate is $\sim\!15\,\mathrm{mm}$,
improving to $\sim\!6\,\mathrm{mm}$ for events with two reconstructed converted photons.
The tracking information from the ID improves the identification of the vertex of the hard interaction, which is needed for the jet selection in the 2-jet category.

With the selection described in Section~\ref{sec:hggselect},
in the diphoton invariant mass range between $100\GeV$ and $160\GeV$,
$23788$ and $35251$ diphoton candidates are observed
in the $7\TeV$ and $8\TeV$ data samples, respectively.

Data-driven techniques~\cite{Collaboration:2011ww} are used to estimate the numbers of $\gamma\gamma$, $\gamma j$ and $jj$ events in the selected sample.
The contribution from the Drell-Yan background is determined from a sample of \ztoee\ decays in data where either one or both electrons pass the photon selection.
The measured composition of the selected sample is approximately $74\%$, $22\%$, $3\%$ and $1\%$ for the $\gamma\gamma$, $\gamma j$, $jj$ and Drell-Yan processes, respectively, demonstrating the dominance of the irreducible diphoton production.
This decomposition is not directly used in the signal search; however, it is used to study the parameterisation of the background modelling.

\subsection{Event categorisation}
\label{sec:hggeventcat}
To increase the sensitivity to a Higgs boson signal, the events are separated
into ten mutually exclusive categories having different mass resolutions and signal-to-background ratios.
An exclusive category of events containing two jets improves the sensitivity to VBF.
The other nine categories are defined by the presence or not of converted photons, $\eta$ of the selected photons, and \ptt,
the component\footnote{$p_{\mathrm{Tt}} = {\left|({{\bf p}_\mathrm{T}^{\gamma_1}} + {{\bf p}_\mathrm{T}^{\gamma_2}}) \times ({{\bf p}_\mathrm{T}^{\gamma_1}} -{{\bf p}_\mathrm{T}^{\gamma_2}})\right|/\left|{{\bf p}_\mathrm{T}^{\gamma_1}} - {{\bf p}_\mathrm{T}^{\gamma_2}}\right|}$,
where ${{\bf p}_\mathrm{T}^{\gamma_1}}$ and ${{\bf p}_\mathrm{T}^{\gamma_2}}$ are the transverse momenta of the two photons.}
of the diphoton \pt\ that is orthogonal to the axis defined by the difference between the two photon momenta~\cite{PTT_OPAL,PTT_ZBoson}.

Jets are reconstructed~\cite{JES} using the anti-$k_t$ algorithm~\cite{Cacciari:2008gp} with radius parameter $R=0.4$.
At least two jets with $\left|\eta\right|<4.5$ and $p_{\mathrm{T}} > 25\,\GeV$ are required in the 2-jet selection.
In the analysis of the $8\TeV$ data, the $p_{\mathrm{T}}$ threshold is raised to $30\GeV$ for jets with $2.5 < |\eta| < 4.5$.
For jets in the ID acceptance ($|\eta|<2.5$),
the fraction of the sum of the \pt\ of tracks, associated with the jet and matched to the selected primary vertex, with respect to the sum of the \pt\ of tracks associated with the jet~(jet vertex fraction, JVF) is required to be at least $0.75$.
This requirement on the JVF reduces the number of jets from proton-proton
interactions not associated with the primary vertex.
Motivated by the VBF topology, three additional cuts are applied in the 2-jet selection:
the difference of the pseudorapidity between the leading and sub-leading jets~(tag jets) is required to be larger than $2.8$,
the invariant mass of the tag jets has to be larger than $400\GeV$, and
the azimuthal angle difference between the diphoton system and the system of the tag jets has to be larger than $2.6$.
About $70\%$ of the signal events in the 2-jet category come from the VBF process.

The other nine categories are defined as follows:
events with two unconverted photons are separated into \emph{unconverted central} ($|\eta|<0.75$ for both candidates) and \emph{unconverted rest} (all other events),
events with at least one converted photon are separated into \emph{converted central} ($|\eta|<0.75$ for both candidates), \emph{converted transition} (at least one photon with $1.3<|\eta|<1.75$) and \emph{converted rest} (all other events).
Except for the \emph{converted transition} category, each category is further divided by a cut at \ptt = 60~GeV into two categories, \emph{low} \ptt\ and \emph{high} \ptt.
MC studies show that signal events, particularly those produced via VBF or associated production ($WH/ZH$ and $t\bar{t}H$),
have on average larger \ptt\ than background events.
The number of data events in each category, as well as 
the sum of all the categories, which is denoted \emph{inclusive}, are given in Table~\ref{tab:SBnumbers}.

\begin{table}[t!]
\begin{center}
\caption{Number of events in the data~($N_\mathrm{D}$) and expected number of signal events~($N_\mathrm{S}$) for $m_H=126.5$~GeV from the \hgg\ analysis, for each category in the mass range $100{\--}160$~GeV.
The mass resolution FWHM (see text) is also given for the 8\,\TeV data.
The Higgs boson production cross section multiplied by the branching ratio into two photons~($\sigma\times B(H\to\gamma\gamma)$) is listed for $m_H=126.5$~GeV.
The statistical uncertainties on $N_\mathrm{S}$ and FWHM are less than 1\,\%.
}
\label{tab:SBnumbers}
\vspace{0.3cm}
\scalebox{0.75}{ 
\begin{tabular}{l|rr|rr|r}
\hline\hline
$\sqrt{s}$ & \multicolumn{2}{c|}{$7\TeV$} & \multicolumn{3}{c}{$8\TeV$} \\
\hline
$\sigma\times B(H\to\gamma\gamma)$ [fb]      &       &  39 &       &  50 & FWHM \\
\cline{1-5}
Category & $N_\mathrm{D}$ & $N_\mathrm{S}$ & $N_\mathrm{D}$ & $N_\mathrm{S}$ & [GeV] \\
\hline
Unconv. central, low \ptt   &  2054 &  10.5 &  2945 &  14.2 & 3.4 \\
Unconv. central, high \ptt  &    97 &   1.5 &   173 &   2.5 & 3.2 \\
Unconv. rest, low \ptt      &  7129 &  21.6 & 12136 &  30.9 & 3.7 \\
Unconv. rest, high \ptt     &   444 &   2.8 &   785 &   5.2 & 3.6 \\
Conv. central, low \ptt     &  1493 &   6.7 &  2015 &   8.9 & 3.9 \\
Conv. central, high \ptt    &    77 &   1.0 &   113 &   1.6 & 3.5 \\
Conv. rest, low \ptt        &  8313 &  21.1 & 11099 &  26.9 & 4.5 \\
Conv. rest, high \ptt       &   501 &   2.7 &   706 &   4.5 & 3.9 \\
Conv. transition            &  3591 &   9.5 &  5140 &  12.8 & 6.1 \\
2-jet                       &    89 &   2.2 &   139 &   3.0 & 3.7 \\
\hline
All categories~(inclusive)  & 23788 &  79.6 & 35251 & 110.5 & 3.9 \\
\hline\hline
\end{tabular}
}
\end{center}
\end{table}

\subsection{Signal modelling}
The description of the Higgs boson signal is obtained from MC, as described in Section~\ref{sec:papersamples}.
The cross sections multiplied by the branching ratio into two photons are given in Table~\ref{tab:SBnumbers} for $m_H=126.5$~GeV.
The number of signal events produced via the ggF process is rescaled to take into account
the expected destructive interference between the $gg\to\gamma\gamma$\; continuum background and ggF~\cite{Dixon:2003yb},
leading to a reduction of the production rate by $2{\--}5\%$ depending on $m_H$ and the event category.
For both the $7\TeV$ and $8\TeV$ MC samples,
the fractions of ggF, VBF, $WH$, $ZH$ and $t\bar{t}H$ production are approximately $88\%$, $7\%$, $3\%$, $2\%$ and $0.5\%$, respectively, for $m_H=126.5$~GeV.

In the simulation, the shower shape distributions are shifted slightly to improve the agreement with the data~\cite{Aad:2010sp},
and the photon energy resolution is broadened~(by approximately $1\%$ in the barrel calorimeter and $1.2{\--}2.1\%$ in the end-cap regions) to account for small differences observed between \ztoee\ data and MC events.
The signal yields expected for the $7\TeV$ and $8\TeV$ data samples are given in Table~\ref{tab:SBnumbers}.
The overall selection efficiency is about $40\%$.

The shape of the invariant mass of the signal in each category is modelled by the sum of a Crystal Ball function~\cite{CSB},
describing the core of the distribution with a width $\sigma_{CB}$,
and a Gaussian contribution describing the tails~(amounting to $<$$10\%$) of the mass distribution.
The expected full-width-at-half-maximum (FWHM) is 3.9~GeV and $\sigma_{CB}$ is 1.6~GeV for the inclusive sample.
The resolution varies with event category (see Table~\ref{tab:SBnumbers}); the FWHM is typically a factor 2.3 larger
than $\sigma_{CB}$.

\subsection{Background modelling}
The background in each category is estimated from data by fitting the
diphoton mass spectrum in the mass range $100{\--}160$~GeV with a
selected model with free parameters of shape and normalisation.
Different models are chosen for the different categories to achieve a
good compromise between limiting the size of a potential bias
while retaining good statistical
power.  A fourth-order Bernstein polynomial function~\cite{Bernstein} is used for the
\emph{unconverted rest} (\emph{low \ptt}), \emph{converted rest} (\emph{low \ptt}) and
\emph{inclusive} categories, an exponential function of a second-order
polynomial for the \emph{unconverted central} (\emph{low \ptt}), \emph{converted central}
(\emph{low \ptt}) and \emph{converted transition} categories, and an exponential
function for all others.

Studies to determine the potential bias have been performed using 
large samples of simulated background events
complemented by data-driven estimates.
The background shapes in the simulation have been cross-checked using
data from control regions.  The potential bias for a given model is
estimated, separately for each category, by performing a maximum likelihood fit
to large samples of simulated background events in the mass range
$100{\--}160$~GeV, of the sum of a signal plus the given
background model.
The signal shape is taken to follow the expectation for a SM Higgs
boson; the signal yield is a free parameter of the fit.
The potential bias is defined by the largest absolute signal yield
obtained from the likelihood fit to the simulated background
samples for hypothesised Higgs boson masses
in the range $110{\--}150$~GeV. 
A pre-selection of background parameterisations is made by requiring that the potential
bias, as defined above, is less than $20\%$ of the statistical uncertainty
on the fitted signal yield.
The pre-selected parameterisation in each category with the best expected sensitivity
for $m_H=125\GeV$ is selected as the background model.

The largest absolute signal yield as defined above
is taken as the systematic uncertainty on the background model.
It amounts to $\pm(0.2{\--}4.6)$ and $\pm(0.3{\--}6.8)$ events, depending on
the category for the $7\TeV$ and $8\TeV$ data samples, respectively.
In the final fit to the data (see Section~\ref{sec:hgg:results})
a signal-like term is included in the likelihood function for each category.
This term incorporates the estimated potential bias, 
thus providing a conservative estimate of
the uncertainty due to the background modelling.

\subsection{Systematic uncertainties}
Hereafter, in cases where two uncertainties are quoted, they refer to the $7\TeV$ and $8\TeV$ data, respectively.
The dominant experimental uncertainty on the signal yield~($\pm8\%$, $\pm11\%$) 
comes from the photon reconstruction and identification efficiency, which is
estimated with data using electrons from $Z$ decays and photons
from $Z\to\ell^+\ell^-\gamma$ events.
Pile-up modelling also affects the expected yields and contributes to the
uncertainty ($\pm4\%$).  Further uncertainties on the signal yield are
related to the trigger ($\pm1\%$), photon isolation ($\pm0.4\%$, $\pm0.5\%$)
and luminosity ($\pm1.8\%$, $\pm3.6\%$).
Uncertainties due to the modelling of the underlying event are $\pm6\%$ for VBF and
$\pm30\%$ for other production processes in the 2-jet category.
Uncertainties on the predicted cross sections and branching ratio are summarised in Section~\ref{sec:CombSyst}.

The uncertainty on the expected fractions of signal events in each category is described in the following.
The uncertainty on the knowledge of the material in front of the calorimeter
is used to derive the amount of possible event migration
between the converted and unconverted categories ($\pm4\%$). 
The uncertainty from pile-up on the population of
the converted and unconverted categories is $\pm2\%$.
The uncertainty from the jet energy scale (JES) amounts to up to
$\pm19\%$ for the 2-jet category,
and up to $\pm4\%$ for the other categories.
Uncertainties from the JVF modelling
are $\pm12\%$~(for the $8\TeV$ data) for the 2-jet category, estimated from $Z+$2-jets events by comparing data and MC.
Different PDFs and scale variations in the 
{\tt HqT}~calculations are used to derive possible event migration
among categories ($\pm9\%$) due to the modelling of the Higgs boson kinematics.

The total uncertainty on the mass resolution is $\pm14\%$.
The dominant contribution ($\pm12\%$) comes from the uncertainty on the energy resolution of the calorimeter,
which is determined from \ztoee\ events.
Smaller contributions
come from the imperfect knowledge of the material in
front of the calorimeter, which affects the extrapolation of the calibration from electrons to
photons ($\pm6\%$), and from pile-up ($\pm4\%$).

\begin{figure}[h!]
\begin{center}
\includegraphics[width=.45\textwidth]{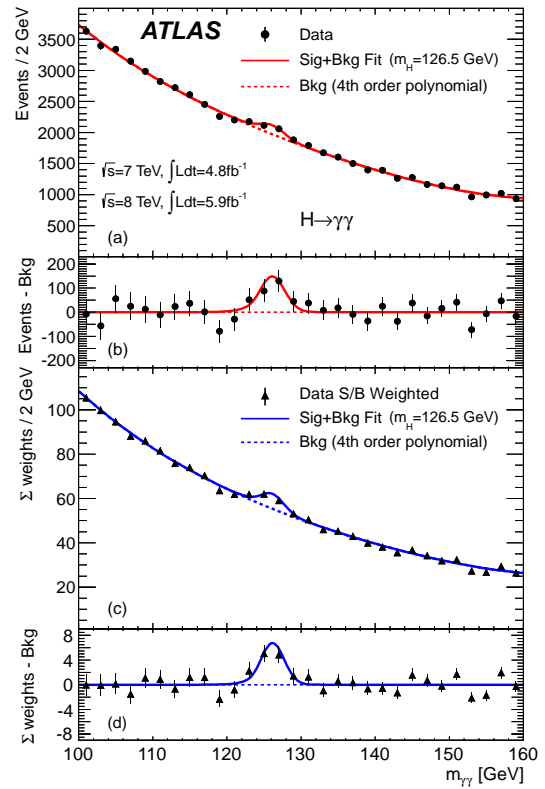}
\end{center}
\caption{The distributions of the invariant mass of diphoton candidates after all selections
  for the combined 7 TeV and 8 TeV data sample.
  The inclusive sample is shown in (a) and a weighted version of the same sample in (c); the
  weights are explained in the text.
  The result of a fit to the data of the sum of a signal component
  fixed to $m_H=126.5\GeV$ and a background
  component described by a fourth-order Bernstein polynomial is superimposed.
  The residuals of the data and weighted data
  with respect to the respective fitted background component are displayed in (b) and (d).
}
\label{fig:fitincl}
\end{figure}

\subsection{Results}
\label{sec:hgg:results}
The distributions of the invariant mass, $m_{\gamma\gamma}$, of the diphoton events, summed over all categories, are shown in Fig.~\ref{fig:fitincl}(a) and (b).
The result of a fit including a signal component fixed to $m_H=126.5\GeV$ and a background component described by a fourth-order Bernstein polynomial is superimposed.

The statistical analysis of the data employs an unbinned likelihood function constructed from
those of the ten categories of the $7\TeV$ and $8\TeV$ data samples.
To demonstrate the sensitivity of this likelihood analysis,
Fig.~\ref{fig:fitincl}(c) and (d) also show the mass spectrum obtained
after weighting events with category-dependent factors reflecting
the signal-to-background ratios. 
The weight $w_i$ for events in category $i\in[1,10]$ for the $7\TeV$ and $8\TeV$ data samples is defined to be $\ln{(1+S_i/B_i)}$, where
$S_i$ is $90\%$ of the expected signal for $m_H=126.5$ GeV,
and $B_i$ is the integral, in a window containing $S_i$, of a background-only fit to the data.
The values $S_i/B_i$ have only a mild dependence on $m_H$.

The statistical interpretation of the excess of events near $m_{\gamma\gamma} = 126.5$ GeV in Fig.~\ref{fig:fitincl} is presented in Section~\ref{sec:Results}.

\section{\texorpdfstring{$\hwwenmun$ channel}{H->WW(*)->evmuv channel}}
\label{sec:hww}

The signature for this channel is two opposite-charge leptons with large transverse
momentum and a large momentum imbalance in the event due to the escaping neutrinos.
The dominant backgrounds are
non-resonant $WW$, $t\bar{t}$, and $Wt$  
production, all of which have real $W$~pairs in the final state.
Other important backgrounds include 
Drell-Yan events ($pp{\to\,}Z/\gamma^{(\ast)}{\to\,}\ell\ell$) with \met{} that may arise from mismeasurement, 
\Wjets\ events in which a jet produces an object reconstructed as the second electron or muon, 
and $W\gamma$ events in which the photon undergoes a conversion.
Boson pair production ($W\gamma^{\ast}/WZ^{(\ast)}$ and $ZZ^{(\ast)}$)  
can also produce opposite-charge lepton pairs with additional leptons that 
are not detected.

The analysis of the 
8 TeV data presented here is 
focused on the mass range $110 < m_H < 200 \GeV$.  
It follows the procedure used for the 7 TeV data, described in Ref.~\cite{ATLAS-4.7fbHWW}, 
except that more stringent criteria are applied to reduce the 
 $W$+jets background and some selections have been modified 
to mitigate the impact of the higher instantaneous luminosity at the LHC in 2012. In particular, 
the higher luminosity results in a larger Drell-Yan background to the same-flavour final states, 
due to the deterioration of the missing transverse momentum resolution.
For this reason, and the fact that the $e\mu$ final state provides more than 85\% of the sensitivity
of the search, the same-flavour final states have not been used in the analysis described here. 

\subsection{Event selection}
\label{sec:selection}

For the 8 TeV \lelm\ search, the data are selected using inclusive single-muon and single-electron triggers.
Both triggers require an isolated lepton with $\pt > 24\GeV$.  
Quality criteria are applied  to suppress
non-collision backgrounds
such as cosmic-ray muons, beam-related
backgrounds, and noise in the calorimeters.
The primary vertex selection follows that described in Section~\ref{sec:h4l}.
Candidates for the $\lelm$ search are pre-selected by requiring exactly two opposite-charge 
leptons of different flavours, with \pt{} thresholds of 25~\GeV{} for the leading lepton and 15~\GeV{} for the
sub-leading lepton. 
Events are classified into two exclusive lepton channels depending on the
flavour of the leading lepton, where  $e\mu$ ($\mu e$) refers to events with a leading electron (muon).  
The dilepton invariant mass is required to be greater than 10~\GeV.

The lepton selection and isolation have more stringent requirements  
than those used for the \htollllp{} analysis (see Section~\ref{sec:h4l}), 
to reduce the larger background from non-prompt leptons
in the $\ell\nu\ell\nu$ final state.  Electron candidates are selected using a combination of 
tracking and calorimetric information~\cite{Aad:2011mk}; the criteria are optimised for 
background rejection, at the expense of some reduced efficiency.  Muon candidates are restricted to those 
with matching MS and ID tracks~\cite{ATLAS-CONF-2011-063}, and therefore are reconstructed over  
$|\eta|< 2.5$.  The isolation criteria require
the scalar sums of the \pt{} of charged particles and of calorimeter topological clusters within 
$\Delta R=\IsoConeSize$ of the lepton direction 
(excluding the lepton itself) each 
to be less than \IsoCutRange\ times the lepton \pt. The exact value differs  
between the criteria for tracks and calorimeter clusters, for both electrons
and muons, and depends on the lepton \pt. 
Jet selections follow those described in Section~\ref{sec:hggeventcat}, except that  
the 
JVF
is required to be
greater than 0.5.

Since two neutrinos are present in the signal final state, 
events are required to have large \met{}. 
${\bf E}_{\rm T}^{\rm miss}$ is the negative vector sum of the transverse momenta 
of the reconstructed objects, including muons, electrons,
photons, jets, and clusters of calorimeter cells not associated with these objects.
The quantity $\metrel$ used in this analysis 
is required  to be greater than \CaloMETCutem~GeV and is defined as:
$\metrel = \met\sin\Delta\phi_{\min}$, where $\Delta\phi_{\min}$ is 
$\min(\Delta\phi, \frac{\pi}{2})$, and \met{} is the magnitude of the vector ${\bf E}_{\rm T}^{\rm miss}$. 
Here, $\Delta\phi$ is the angle between 
${\bf E}_{\rm T}^{\rm miss}$ and the transverse momentum of the 
nearest lepton or jet with $\pt > 25~\GeV$.
Compared to \met, \metrel\ has increased rejection power for events in which the \met\ 
is generated by a neutrino in a jet or the mismeasurement of an object, since in
those events the ${\bf E}_{\rm T}^{\rm miss}$ tends to point in the direction of the object.
After the lepton isolation and \metrel\ requirements that define the pre-selected sample, 
the multijet background is negligible and the Drell-Yan background is much reduced. 
The Drell-Yan contribution becomes very small after the topological selections, described below, 
are applied.

The background rate and composition depend significantly on the jet multiplicity, as
does the signal topology. 
Without accompanying jets, the signal originates almost entirely from the ggF
process and the background is dominated by $WW$ events.  
In contrast, when produced in association with two or more jets, the signal
contains a much larger contribution from the VBF process compared to the ggF process, and the background is
dominated by \ttbar{} production.
Therefore, to maximise the sensitivity to SM Higgs events, further selection criteria depending on
the jet multiplicity are applied to the pre-selected sample.
The data are subdivided into 
\ZeroJet, \OneJet{} and \TwoJet{} search channels according to the number of jets in the final state, 
with the \TwoJet{} channel also including higher jet multiplicities.

Owing to spin correlations in the $WW^{(\ast)}$ system arising from the spin-0 nature of
the SM Higgs boson and the V-A structure of the $W$ boson decay vertex, the charged leptons 
tend to emerge from the primary vertex pointing
 in the same direction~\cite{pr_55_167}.  This kinematic feature is
exploited for all jet multiplicities by requiring that 
$|\Delta\phi_{\ell\ell}| < 1.8$, 
and the dilepton invariant mass, $\hwwmll$, be less than 50~\GeV{}
for the \ZeroJet{} and \OneJet{} channels. For the \TwoJet{} channel, the $\hwwmll$
upper bound is increased to 80~\GeV{}.

In the \ZeroJet{} channel, the magnitude $\ptll$ of the transverse momentum of the 
dilepton system, $\vpTll = {\bf p}_{\rm T}^{\ell 1} + {\bf p}_{\rm T}^{\ell 2}$,
is required to be greater than 30~\GeV{}. This improves the rejection of the Drell-Yan background.

In the \OneJet{} channel, backgrounds from top quark production are suppressed by
rejecting events containing a $b$-tagged jet, as determined using a
$b$-tagging algorithm that uses a neural network
and exploits the topology of weak decays of $b$- and $c$-hadrons~\cite{ATLAS-btag-algs}. 
The total transverse momentum, $\pt^{\rm tot}$,
defined as the magnitude of the vector sum
${\bf p}_{\rm T}^{\rm tot}={\bf p}_{\rm T}^{\ell 1} +
{\bf p}_{\rm T}^{\ell 2}+{\bf p}_{\rm T}^{j}+{\bf E}_{\rm T}^{\rm miss}$,
is required to be smaller than 30~\GeV{} to suppress top background events  
that have jets with $\pt$ below the threshold defined for jet counting. In order to reject the 
background from $Z{\rightarrow\,}\tau\tau$, 
the $\tau\tau$ invariant mass, $m_{\tau\tau}$, is computed under the assumptions 
that the reconstructed leptons are $\tau$ lepton decay products. In addition  the neutrinos
produced in these decays are assumed to be the only source of \met{} and to be 
collinear with the leptons~\cite{CollApp}.
Events with $|m_{\tau\tau}-m_{Z}|<25~\GeV$ are rejected
if the collinear approximation yields a physical solution.

The \TwoJet{} selection 
follows the \OneJet{} selection described above, with the $\pt^{\rm tot}$
definition modified to include all selected jets. Motivated by the VBF topology,
several additional criteria are applied to the tag jets, defined as
the two highest-\pt{} jets in the event.
These are required to be separated in rapidity by a distance $|\Delta y_{jj}| > 3.8$ and to have 
an invariant mass, $m_{jj}$, larger than 500~\GeV. 
Events with an additional jet with  $\pt > 20~\GeV$  between the tag jets ($y_{j1} < y < y_{j2}$) 
are rejected. 

A transverse mass variable, \mT~\cite{Barr:2009mx}, is used to test for the 
presence of a signal for all jet multiplicities. This variable is defined as: 
\begin{displaymath}
  \label{eq:mT}
  \mT = \sqrt{(E_{\rm T}^{\ell\ell}+\met)^{2} - |\vpTll+{\bf E}_{\rm T}^{\rm miss}|^{2}},
\end{displaymath}
where $E_{\rm T}^{\ell\ell} = \sqrt{|\vpTll|^{2}+m_{\ell\ell}^{2}}$.  
The statistical analysis of the data 
uses a fit to the $\mT$ distribution in the signal 
region after the $\Delta\phi_{\ell\ell}$
requirement (see Section~\ref{sec:hwwresults}), which results in increased sensitivity compared to the 
analysis described in Ref.~\cite{ATLAS-2fbHWW}. 

 For a SM Higgs boson with $m_H = 125$\,\GeV, the cross section times branching ratio to 
the $e\nu\mu\nu$ final state is 88\,fb for $\sqrt{s}=7$\,TeV, increasing to 112 fb at $\sqrt{s}=8$\,TeV. 
The combined acceptance times efficiency of the 8\,TeV \ZeroJet{} and \OneJet{} selection relative to the  
ggF production cross section times branching ratio is about 7.4\%.  The acceptance times efficiency of the 
8\,TeV \TwoJet{} selection relative to the VBF production cross section times branching ratio is about 14\%.  
Both of these figures are based on the number of events selected before the final $m_{\rm T}$ criterion is applied
(as described in Section~\ref{sec:hwwresults}).

\subsection{Background normalisation and control samples}
\label{sec:control}

The leading backgrounds from SM			
processes producing two isolated high-$\pt$ leptons are $WW$ and top (in this section, 
``top'' background always includes both $t\bar{t}$ and single top, unless otherwise noted).
These are estimated using partially data-driven techniques based on normalising the MC
predictions to the data in control regions dominated by the relevant background source.
The \Wjets{} background
is estimated from data for all jet multiplicities.
Only the small backgrounds from Drell-Yan and diboson processes other than $WW$, 	
as well as the $WW$ background for the \TwoJet{} analysis, are estimated using MC simulation.  

The control and validation regions are defined by selections similar to those used for the signal region
but with some criteria reversed or modified to obtain signal-depleted samples enriched
in a particular background.  
The term ``validation region'' distinguishes these regions from the control
regions that are used to directly normalise the backgrounds.
Some control regions have significant contributions from backgrounds other than the targeted
one, which introduces dependencies among the background estimates.  These correlations
are fully incorporated in the fit to the $m_{\rm T}$ distribution.  In the following sections, 
each background estimate is described after any others on which it depends.
Hence, the largest background ($WW$) is described last.

\subsubsection{\Wjets{} background estimation} 
\label{sec:wjets-control}

The \Wjets{} background contribution is estimated using a control 
sample of events where one of the two leptons satisfies the identification and
isolation criteria described in Section~\ref{sec:selection}, and the other
lepton fails these criteria but satisfies a
loosened selection  (denoted ``anti-identified''). Otherwise, events in this sample are required to pass all the 
signal selections.  The dominant contribution to this sample comes from 
\Wjets{} events in which a jet produces an object that is reconstructed as 
a lepton.   This object may be either a true electron or muon from 
the decay of a heavy quark, or else a product of the fragmentation identified 
as a lepton candidate. 

The contamination in the signal region is obtained by scaling the number of 
events in the data control sample by a transfer factor.  
The transfer factor is defined here as the ratio of the number of  identified lepton 
candidates passing all selections to the number of anti-identified leptons.  
It is calculated as a function of the anti-identified lepton $\pt$ using
a data sample dominated by QCD jet production (dijet sample) 
 after subtracting the residual contributions from 
leptons produced by leptonic $W$ and $Z$ decays, as estimated from data.   
The small remaining lepton contamination, which includes $W\gamma^{(\ast)}/WZ^{(\ast)}$ events, 
is subtracted using MC simulation.   

\begin{figure}[hbt!]
\begin{center}
\subfigure[]{\includegraphics[width=0.45\textwidth]{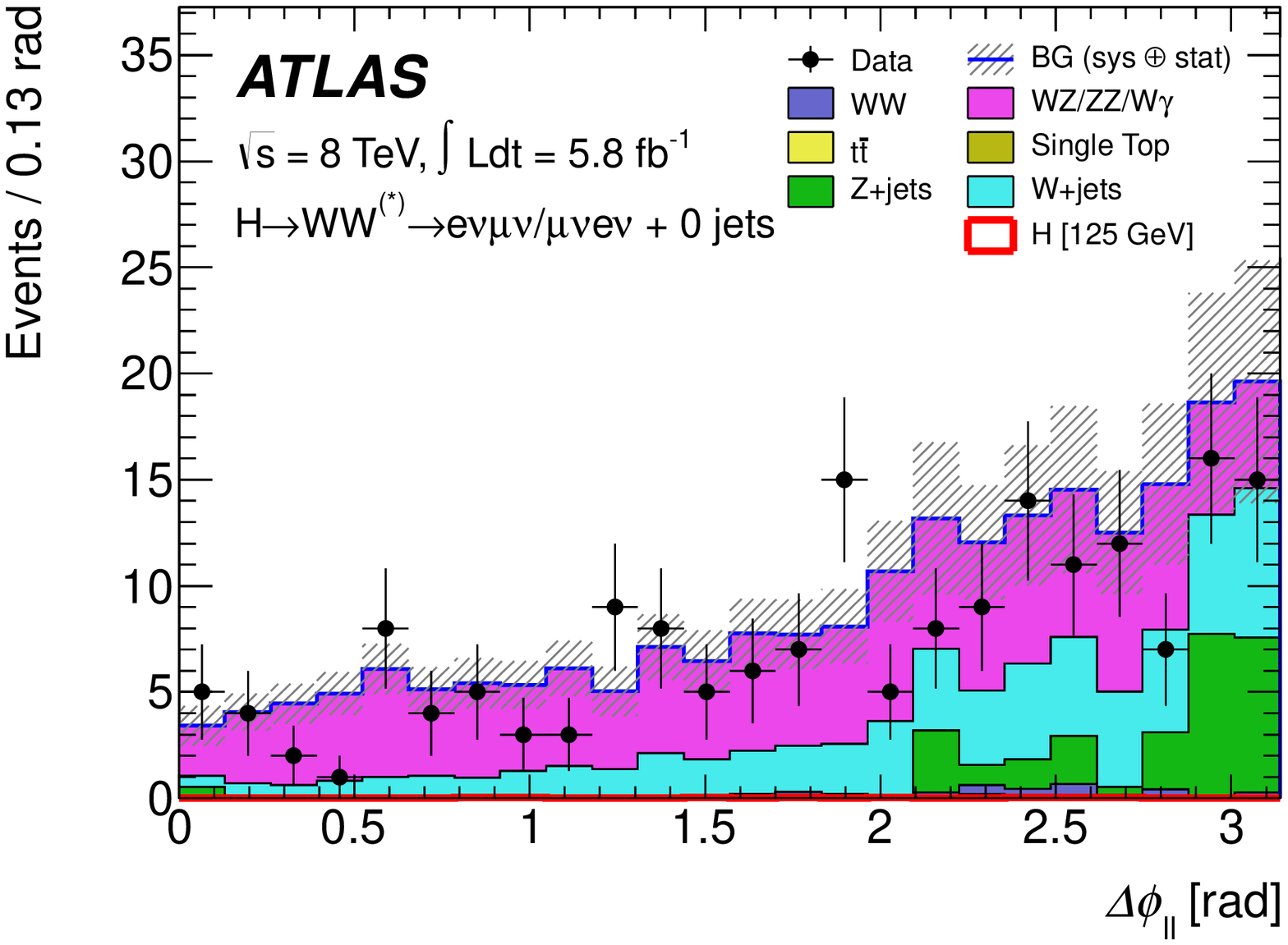} }
\subfigure[]{\includegraphics[width=0.45\textwidth]{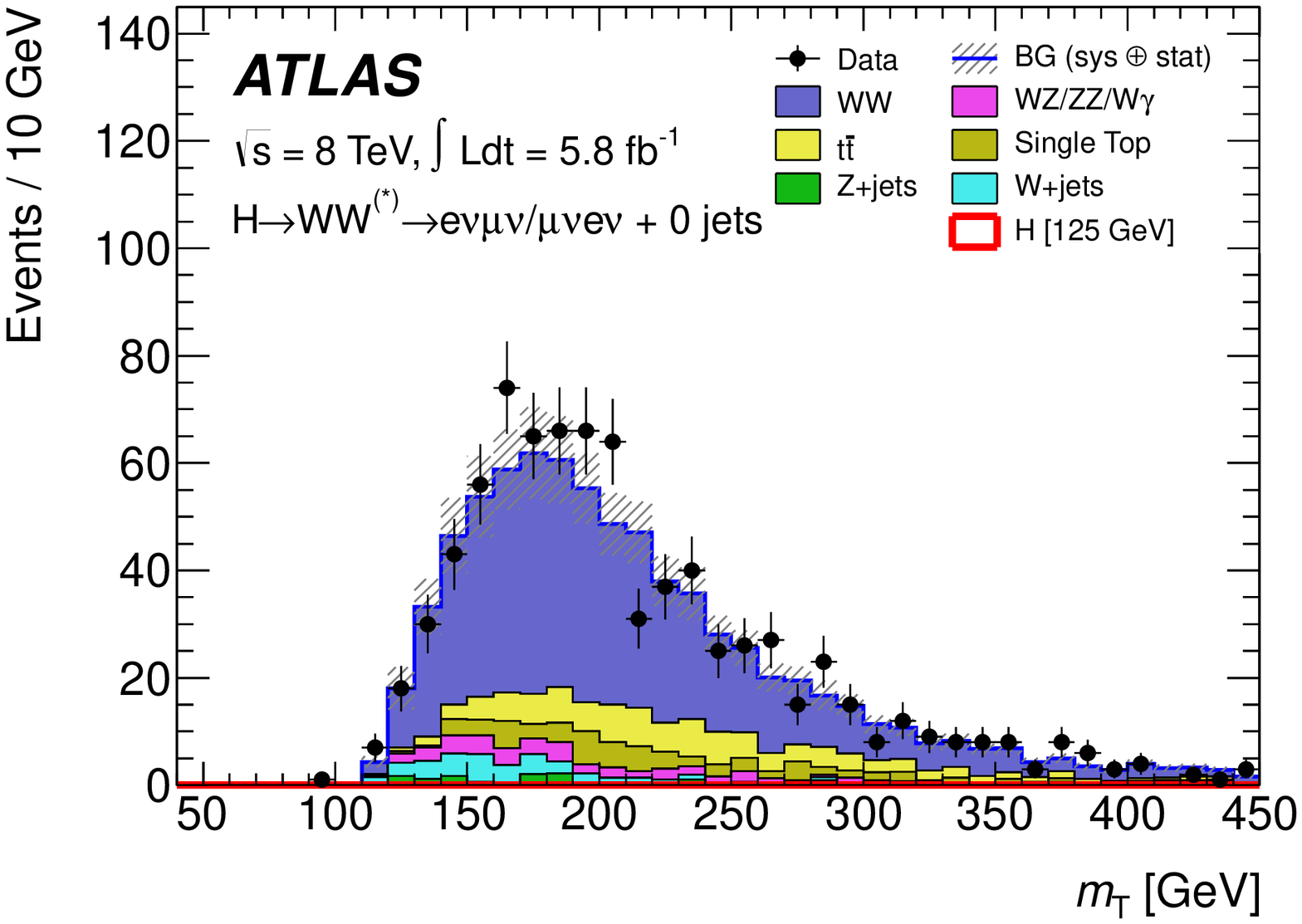}}
\caption{Validation and control distributions for the \lelm\ analysis. 
a) $\Delta\phi_{\ell\ell}$ distribution in the 
     same-charge validation region after the \metrel\ and zero-jet requirements.
b) $m_{\rm T}$ distribution in the $WW$ control region for the  \ZeroJet{} channel.  
    The $e\mu$ and $\mu e$ final states are combined. 
    The hashed area indicates the total uncertainty on the background prediction.
The expected signal for $m_{H} = 125\GeV$ is negligible 
and therefore not visible.
  }
\label{fig:WWCRplots}
\end{center}
\end{figure}

The processes producing the majority of same-charge dilepton events,
\Wjets, $W\gamma^{(\ast)}/WZ^{(\ast)}$ and $Z^{(\ast)}Z^{(\ast)}$, are
all backgrounds in the opposite-charge signal region. 
\Wjets{} and \Wg{} backgrounds are particularly important in a search optimised for a low Higgs
boson mass hypothesis.  Therefore, the normalisation and kinematic features of 
same-charge dilepton events are used to validate the predictions of these backgrounds.
The predicted number of same-charge events 
after the \metrel\ and zero-jet requirements
is 216 $\pm$ 7 (stat) $\pm$ 42 (syst), while 182 events are observed in the data. 
Satisfactory agreement between data and simulation is observed in various kinematic distributions, including those of 
$\Delta\phi_{\ell\ell}$ (see Fig.~\ref{fig:WWCRplots}(a)) and the transverse mass. 

\subsubsection{Top control sample}
\label{sec:top-control}

In the \ZeroJet{} channel, the top quark background prediction is first normalised using
events satisfying the pre-selection criteria described in Section~\ref{sec:selection}. 
This sample is selected without jet multiplicity or $b$-tagging requirements, and the majority of
events contain top quarks.  Non-top contributions are subtracted using predictions
from simulation, except for \Wjets, which is estimated using data.  After this normalisation is performed, 
the fraction of events with zero jets that pass all selections is evaluated. This fraction is small (about 3\%),
since the top quark decay $\ttoWb$ has a branching ratio of nearly 1.
Predictions of this fraction from MC simulation are sensitive to theoretical uncertainties such 
as the modelling of initial- and final-state radiation, 
as well as experimental uncertainties, especially that on the jet energy scale. 
To reduce the impact of these uncertainties,  the top quark background determination uses 
data from a $b$-tagged control region in which the one-to-two jet ratio 
is compared to the MC simulation ~\cite{Aad:2011qi}.
The resulting correction factor to a purely MC-based background estimate after all selections amounts to
$1.11 \pm 0.06$ (stat).

In the \OneJet{} and \TwoJet{} analyses, the top quark background predictions are 
normalised to the data using control samples defined by reversing the $b$-jet
veto and removing the requirements on $\Delta\phi_{\ell\ell}$ and $m_{\ell\ell}$.
The $|\Delta y_{\rm jj}|$ and $m_{\rm jj}$ requirements are included in the
definition of the 2-jet control region.
The resulting samples are dominated by top quark events.  
The small contributions from other sources are taken into account   
using MC simulation and the data-driven \Wjets\ estimate.
Good agreement between data and MC simulation is observed for the total numbers of events and the 
shapes of the $m_{\rm T}$ distributions.
The resulting normalisation factors are \TopSFonejet\ for the \OneJet{} control region and 
\TopSFtwojet\ for the \TwoJet{} control region. Only the statistical uncertainties are quoted.

\subsubsection{$WW$ control sample}
\label{sec:WW-control}

The MC predictions of the $WW$ background in the \ZeroJet{} and \OneJet{} analyses,
summed over lepton flavours, are normalised using control regions defined with
the same selections as for the signal region except that the
$\Delta\phi_{\ell\ell}$ requirement is removed and the upper
bound on \hwwmll\ is replaced with a lower bound: $\hwwmll > 80~\GeV$.
The numbers of events and the shape of the $m_{\rm T}$ distribution in the control regions 
are in good agreement between data and MC, as shown in Fig.~\ref{fig:WWCRplots}(b).
$WW$ production contributes  about 70\% of the  
events in the 0-jet control region and about 45\% in the 1-jet region.
Contaminations from sources other than $WW$ are derived as for the signal
region, including the data-driven \Wjets{} and top estimates.
The resulting normalisation factors with their associated statistical uncertainties 
are \WWSFzerojet\ for the \ZeroJet{} control region and 
\WWSFonejet\ for the \OneJet{} control region.

\subsection{Systematic uncertainties}
\label{sec:systematics}

The systematic uncertainties that have the largest impact on the sensitivity of the search are the 
theoretical uncertainties associated with the signal. These are described 
in Section~\ref{sec:Results}.  The main experimental uncertainties are associated with the JES, 
the jet energy resolution (JER), pile-up, \met{}, 
the $b$-tagging efficiency, the $W$+jets transfer factor, and the integrated luminosity. 
The largest uncertainties on the backgrounds 
include $WW$ normalisation and modelling, top normalisation, 
and $W\gamma^{(\ast)}$ normalisation. The \TwoJet{} systematic uncertainties are dominated by 
the statistical uncertainties in the data and the MC simulation, and are therefore not discussed further. 

Variations of the jet energy scale within the systematic uncertainties can cause events to migrate between the jet bins.
The uncertainty on the JES varies from $\pm 2$\% to $\pm 9$\% as a function of
jet $\pt$ and $\eta$ for jets with  $\pt > 25\GeV$ and $|\eta| < 4.5$~\cite{JES}. 
The largest impact of this uncertainty on the total signal (background) yield amounts to 7\% (4\%) in the 0-jet (1-jet) bin.
The uncertainty on the JER is estimated from \emph{in situ} measurements and it impacts
mostly the \OneJet{} channel, where its effect on the total signal and background yields is 4\% and 2\%, respectively.
An additional contribution to the JES uncertainty arises from pile-up, and is estimated to vary between $\pm 1$\%
and $\pm 5$\% for multiple $pp$ collisions in the same bunch crossing 
and up to $\pm 10$\% for neighbouring bunch crossings. This uncertainty affects mainly the 
\OneJet{} channel, where its impact on the signal and background yields is 4\%  and 2\%, respectively.
JES and lepton momentum scale uncertainties are propagated to the
\met{} measurement. Additional contributions to the \met{} uncertainties arise from jets with $\pt < 20\GeV$
and from low-energy calorimeter deposits not associated with reconstructed
physics objects~\cite{MET}. 
The impact of the \met{} uncertainty on the total signal and background yields is $\sim$3\%.
The efficiency of the $b$-tagging algorithm is calibrated
using samples containing muons reconstructed in the vicinity of
jets~\cite{ATLAS-CONF-2012-043}. The uncertainty on the $b$-jet tagging
efficiency varies between $\pm 5$\% and $\pm 18$\%  as a function of the jet \pt{}, and its 
impact on the total background yield is 10\% for the \OneJet{} channel. 
The uncertainty in the $W$+jets transfer factor is dominated by differences in jet properties between dijet and
$W$+jets events as observed in MC simulations.
The total uncertainty on this background is approximately $\pm \WjetsSRErrorEle$, resulting in an uncertainty on  
the total background yield of 5\%.
The uncertainty on the integrated luminosity is $\pm 3.6$\%.		

A fit to the distribution of \mT{} is performed in order to obtain 
the signal yield for each mass hypothesis (see Section~\ref{sec:hwwresults}). 
Most theoretical and experimental uncertainties do not
produce statistically significant changes
to the \mT{} distribution.  The uncertainties
that do produce significant changes of the distribution of \mT{}
have no appreciable effect on the final results, with the exception 
of those associated with the $WW$ background. In this case, an uncertainty is included to take into account 
differences in the distribution of \mT\ and normalisation observed between the MCFM \cite{Campbell:2011bn}, MC@NLO+HERWIG and POWHEG+PYTHIA generators.
The potential impact of interference between resonant (Higgs-mediated) and non-resonant $\ggWW$ diagrams~\cite{Campbell:2011cu} 
for $\mT > m_H$ was investigated and found to be negligible.
The effect of the $WW$ normalisation, modelling, and shape systematics on the total background yield 
is 9\% for the \ZeroJet{} channel and 19\% for the \OneJet{} channel. 
The uncertainty on the shape of the total background is dominated by the
uncertainties on the normalisations of the individual backgrounds.
The main uncertainties on the top background in the \ZeroJet{} analysis include those associated with 
interference effects between $t\bar{t}$ and single top, initial state an final state radiation, 
$b$-tagging, and JER. The impact on the total background yield in the 0-jet bin is 3\%.  
For the \OneJet{} analysis, the impact of the top background on the total yield is 14\%. 
Theoretical uncertainties on the $W\gamma$ background normalisation are evaluated for each jet
bin using the procedure described in Ref.~\cite{Stewart:2011cf}. They are 
$\pm 11$\% for the 0-jet bin and $\pm 50$\% for the 1-jet bin. 
For $W\gamma^{\ast}$ with $m_{\ell\ell} < 7$~\GeV, a k-factor of $1.3\pm 0.3$ is applied to 
the MadGraph LO prediction based on the comparison with the MCFM NLO calculation.
The k-factor for $W\gamma^{\ast}/WZ^{(\ast)}$ with $m_{\ell\ell} > 7$~\GeV\ is 1.5 $\pm$ 0.5.
These uncertainties affect mostly the \OneJet{} channel, where their impact 
on the total background yield is approximately 4\%.


\begin{table}[!htbp]
\centering
\caption{
The expected numbers of signal ($m_{H}=125\GeV$) and background
events after all selections, 
including a cut on the transverse mass of 
$0.75\,m_{H}<m_{\rm T}<m_{H}$ for $m_{H}=125\GeV$. 
The observed numbers of events in data are also displayed.
The $e\mu$ and $\mu e$ channels are combined. 
The uncertainties shown are the combination of the statistical and all
systematic uncertainties, taking into account the constraints from control
samples. 
  For the \TwoJet\ analysis, backgrounds
with fewer than 0.01 expected events are marked with `-'.
}
\label{hww_cutflow}
\vspace*{0.2cm}
\scalebox{0.8}{
\begin{tabular}{l|r@{$\,\pm \,$}l r@{$\,\pm \,$}l r@{$\,\pm \,$}l}
\hline
\hline
  & \multicolumn{2}{c}{\ZeroJet} & \multicolumn{2}{c}{\OneJet} &  \multicolumn{2}{c}{\TwoJet}  \\
\hline
 Signal & 20 & 4 & 5 & 2 & 0.34 & 0.07 \\ \hline
 $WW$ & 101 & 13 & 12 & 5 & 0.10 & 0.14 \\ 
 $WZ^{(\ast)}/ZZ/W\gamma^{(\ast)}$ & 12 & 3 & 1.9 & 1.1 & 0.10 & 0.10 \\ 
 $t\bar{t}$ & 8 & 2 & 6 & 2 & 0.15 & 0.10 \\ 
$tW/tb/tqb$  & 3.4 & 1.5 & 3.7 & 1.6 & \multicolumn{2}{c}{-} \\ 
 $Z/\gamma^{\ast}+\mathrm{jets}$ & 1.9 & 1.3 & 0.10 & 0.10 &\multicolumn{2}{c}{-} \\ 
 $W+\mathrm{jets}$ & 15 & 7 & 2 & 1 & \multicolumn{2}{c}{-} \\ \hline
 Total Background & 142 & 16 & 26 & 6 & 0.35 & 0.18 \\ \hline
 Observed  & \multicolumn{2}{c}{185} & \multicolumn{2}{c}{38} &  \multicolumn{2}{c}{0}  \\
\hline
\hline
\end{tabular}
}
\end{table}

\subsection{Results}
\label{sec:hwwresults}

Table~\ref{hww_cutflow} shows the numbers of events expected from a SM Higgs boson with $m_{H}=125\GeV$ and from the backgrounds, as well as
the numbers of candidates observed in data, after application of all selection criteria plus an
additional cut on $\mT$ of  $0.75\,m_{H} < m_{\rm T} < m_{H}$. 
The uncertainties shown in Table~\ref{hww_cutflow} include 
the systematic uncertainties discussed in
Section~\ref{sec:systematics}, constrained by the use of the control
regions discussed in Section~\ref{sec:control}.  
An excess of events relative to the background expectation is observed in the data.

\begin{figure}[hbtp]
  \centering
  \includegraphics[width=0.47\textwidth]{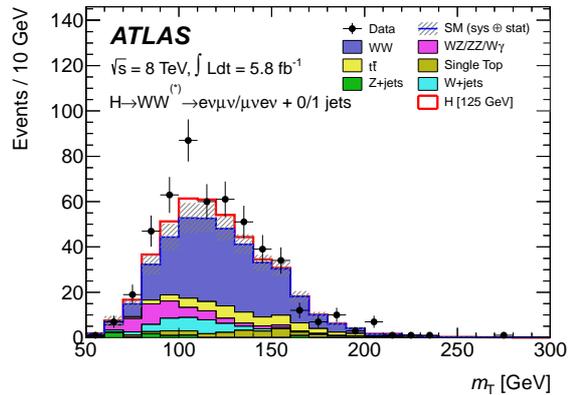}
  \vspace*{-0.5cm}
  \caption{ Distribution of the transverse mass, \mT, in the \ZeroJet{} and \OneJet{}
    analyses with both $e \mu$ and $\mu e$ channels combined, for events satisfying all selection criteria.
The expected signal for $m_{H} = 125\GeV$  is shown stacked on top of the background prediction. 
 The $W$+jets background is estimated  from data, and  
    $WW$ and top background MC predictions are normalised to the data using control regions. 
    The hashed area indicates the total uncertainty on the background prediction.
    }
  \label{fig:papermT}
\end{figure}

 Figure~\ref{fig:papermT} shows the distribution of the transverse mass after all
selection criteria in the \ZeroJet{} and \OneJet{} channels combined, and for both  lepton
channels together.  

The statistical analysis of the data employs a binned likelihood function
constructed as the product of Poisson
probability terms for the $e\mu$ channel and the $\mu e$ channel. 
The mass-dependent cuts on \mT{} described above are not used.
Instead, the \ZeroJet{} (\OneJet) signal regions are subdivided into five (three)
\mT{} bins. For the \TwoJet{} signal region, only 
the results integrated over \mT{} are
used, due to the small number of events in the final sample. 
The statistical interpretation of the observed excess of events is presented in 
Section~\ref{sec:Results}.

\section{Statistical procedure\label{sec:statproc}}

The statistical procedure used to interpret the data is
described in
Refs.~\cite{paper2012prd,LHC-HCG,Moneta:2010pm,HistFactory,ROOFIT}.
The parameter of interest is the global signal strength factor $\mu$,
which acts as a scale factor on the total number of events predicted
by the Standard Model for the Higgs boson signal.  This factor is 
defined such that $\mu=0$ corresponds to the background-only hypothesis
and $\mu=1$ corresponds to the SM Higgs boson signal in addition to
the background.  Hypothesised values of $\mu$ are tested with a
statistic $\lambda(\mu)$ based on the profile likelihood
ratio~\cite{Cowan:2010st}.  This test statistic extracts the
information on the signal strength from a full likelihood fit to the data. The
likelihood function includes
all the parameters that describe the systematic uncertainties and their
correlations.

Exclusion limits are based on the $CL_s$
prescription~\cite{Read:2002hq}; a value of $\mu$ is regarded as
excluded at 95\%~CL when $CL_s$ is less than 5\%.  A SM Higgs boson
with mass $\mh$ is considered excluded at 95\% confidence level (CL) when $\mu=1$ is
excluded at that mass.  The significance of an excess in the data is
first quantified with the local $p_0$, 
the probability that the background can produce a fluctuation greater than or
equal to the excess observed in data.
The equivalent formulation in terms of number of standard
deviations, $Z_l$, is referred to as the local significance. 
%
%
The global probability for the most significant excess to be observed 
anywhere in a given search region  is estimated with the method described in 
Ref.~\cite{Gross:2010qma}. The ratio of the global to the local probabilities, 
the trials factor used to correct for the "look elsewhere" effect, increases 
with the range of Higgs boson mass hypotheses considered, the mass resolutions 
of the channels involved in the combination, and the significance of the excess.

The statistical tests are
performed in steps of values of the hypothesised Higgs boson mass
$m_H$.
The asymptotic approximation~\cite{Cowan:2010st} upon which the results are
based has been validated with the method described in Ref.~\cite{paper2012prd}.

The combination of individual search sub-channels for a specific Higgs
boson decay, and the full combination of all search channels, are based
on the global signal strength factor $\mu$ and on the identification of
the nuisance parameters that correspond to the correlated sources of
systematic uncertainty described in Section~\ref{sec:CombSyst}.


\begin{table*}[htb]
\center
\caption{Summary of the individual channels entering  the  combination. 
The transition points between separately optimised  \mh\ regions are indicated where applicable. In channels sensitive
to associated production of the Higgs boson, $V$ indicates a $W$ or $Z$ boson. The symbols $\otimes$ and $\oplus$ represent direct products and sums over sets of selection requirements, respectively. 
\label{tab:channels}} 
\vspace{0.cm}
\footnotesize
\mbox{ \hspace{-.35cm}
\begin{tabular}{c|ccccc}\hline\hline
  Higgs Boson	                & Subsequent  &
  \multirow{2}{*}{Sub-Channels}	&$\mh$ Range  & {$\int$ L $dt$ } & \multirow{2}{*}{Ref.} \\ 
  Decay &  Decay 		&   &  [\GeV] & [\infb]   \\ \hline\hline
  \multicolumn{5}{c}{2011 $\sqrt{s}=$7 TeV} \\ \hline\hline

  \multirow{3}{*}{$H\to ZZ^{(*)}$} 	& $4\ell$ & $\{4e,2e2\mu,2\mu2e,4 \mu\}$ & 110--600 & 4.8 & \cite{ATLAS-CONF-2012-092}  \\
  & $\ell\ell\nu\bar{\nu}$ & $\{ee,\mu\mu\}$ $\otimes$ \{low, high pile-up\} & 200--280--600 & 4.7 & \cite{may_llvv}  \\
  & $\ell\ell q\bar{q}$ &  \{$b$-tagged, untagged\} & 200--300--600 & 4.7 & \cite{may_llqq}   \\ \hline

  \multirow{1}{*}{$H\to\gamma\gamma$} & -- & 10 categories \{\ptt\ $\otimes$ $\eta_\gamma$ $\otimes$ $\rm conversion$\} $\oplus$ \{2-jet\}  & 110--150 & 4.8 &  \cite{ATLAS-CONF-2012-091} \\\hline

  \multirow{2}{*}{$H\to WW^{(*)}$} 	& $\ell\nu\ell\nu$ &  $\{ee,e\mu/\mu e,\mu\mu\}$ $\otimes$ \{0-jet, 1-jet, 2-jet\} $\otimes$  \{low, high pile-up\}  & 110--200--300--600 & 4.7 & \cite{ATLAS-4.7fbHWW} \\
  & $\ell\nu q{q}'$ & $\{e,\mu\}$ $\otimes$ \{0-jet, 1-jet, 2-jet\} & 300--600 & 4.7 & \cite{may_lvqq}  \\ \hline
  \multirow{4}{*}{$H\to \tau\tau$} 	& $\tau_{\rm lep}\tau_{\rm lep}$ &  $\{e\mu\} \otimes \{$0-jet$\}$  $\oplus$ \{$\ell\ell$\} $\otimes$ \{1-jet, 2-jet, $VH\}$ & 110--150 & 4.7  & \\
  & \multirow{2}{*}{$\tau_{\rm lep} \tau_{\rm had}$} & $\{e,\mu \}$ $\otimes$ \{0-jet\} $\otimes$  $\{ E_{\rm T}^{\rm miss}  < 20~\textrm{\GeV}, E_{\rm T}^{\rm miss}  \ge 20~\textrm{\GeV}\} $  &  \multirow{2}{*}{110--150} &  \multirow{2}{*}{4.7} &  \cite{may_tautau}   \\
  & & $\oplus$  $\{e,\mu \}$ $\otimes$ \{1-jet\} $\oplus$ \{$\ell$\} $\otimes$ \{2-jet\}  &  &  &  \\
  & $\tau_{\rm had}\tau_{\rm had}$ & \{1-jet\} & 110--150  & 4.7 &  \\ \hline
  \multirow{3}{*}{$VH\to Vb{b}$} & $Z\to\nu {\nu}$ & $ E_{\rm T}^{\rm miss}\in \{120-160, 160-200, \ge 200$ \GeV \}   & 110--130  & 4.6 & \\
  & $W\to\ell\nu$ & $p_{\rm T}^W\in \{< 50, 50-100, 100-200, \ge 200$  \GeV \} & 110--130 & 4.7  & \cite{may_bb}\\
  & $Z\to \ell\ell$ &   $p_{\rm T}^Z\in \{< 50, 50-100, 100-200, \ge 200$  \GeV \}  & 110--130 & 4.7  &\\ \hline \hline
  \multicolumn{5}{c}{2012 $\sqrt{s}=$8 TeV} \\ \hline\hline
  \multirow{1}{*}{$H\to ZZ^{(*)}$} 	& $4\ell$ & $\{4e,2e2\mu,2\mu2e,4 \mu\}$ & 110--600 & 5.8  & \cite{ATLAS-CONF-2012-092}  \\\hline
  \multirow{1}{*}{$H\to\gamma\gamma$} & -- & 10 categories \{\ptt\ $\otimes$ $\eta_\gamma$ $\otimes$ $\rm conversion$\} $\oplus$ \{2-jet\}  & 110--150 & 5.9  & \cite{ATLAS-CONF-2012-091} \\\hline
  \multirow{1}{*}{$H\to WW^{(*)}$} 	& $e\nu\mu\nu$ &  $\{e\mu,\mu e\}$ $\otimes$ \{0-jet, 1-jet, 2-jet\}  & 110--200 & 5.8 & \cite{ATLAS-CONF-2012-098} \\ \hline \hline
\end{tabular}}
\end{table*}

\begin{table*}[!htb]
   \centering
   \caption{Characterisation of the excess in the \htollllp, \hgg\ and \hwwlnln\ channels 
        and the combination of all channels listed in Table~\ref{tab:channels}.
        The  mass value $m_{\rm max}$ for which the
        local significance is maximum, the maximum observed local significance $Z_l$ and the expected local significance $E(Z_l)$ 
        in the presence of a SM Higgs boson signal at $m_{max}$ are given.  
        The best fit value of the signal strength  
	parameter $\hat\mu$ at $\mh=126$~\GeV\ is shown with the total uncertainty.
	The expected and observed mass ranges excluded at 95\%\ CL (99\%\ CL, indicated by a~*) are also given,
        for the combined $\sqrt{s}=7$\,TeV and $\sqrt{s}=8$\,TeV data.
     \label{tab:statsummary}}
    \vspace{0.3cm}
    \resizebox{\textwidth}{!}{
    \begin{tabular}{cc|ccc|c|cc}
      \hline\hline
Search channel & Dataset & $m_{\rm max}$ [GeV] &  $Z_l\,[\sigma]$   & $E(Z_l)\,[\sigma]$ & $\hat{\mu}(\mh=126\,\GeV)$ & Expected exclusion [GeV] & Observed exclusion [GeV] \\
      \hline
       \multirow{3}{*}{\htollllp} & 7\,TeV   & 125.0   & 2.5 & 1.6 & $1.4\pm 1.1$  \\
                                  & 8\,TeV   & 125.5   & 2.6 & 2.1 & $1.1\pm 0.8$  \\
                                  & 7 \& 8\,TeV & 125.0& 3.6 & 2.7 & $1.2\pm 0.6$& \excludedrangeexpaBrief,  \excludedrangeexpbBrief & \excludedrangeaBrief, \excludedrangebBrief \\
      \hline
      \multirow{3}{*}{\hgg}  & 7\,TeV      & 126.0      & 3.4 & 1.6  & $2.2\pm 0.7$\\
                             & 8\,TeV      & 127.0      & 3.2 & 1.9  & $1.5\pm 0.6$     \\
                             & 7 \& 8\,TeV & 126.5      & 4.5 & 2.5  & $1.8\pm 0.5$& 110--140 &  112--123, 132--143  \\
      \hline
      \multirow{3}{*}{\hwwlnln} & 7\,TeV      & 135.0  & 1.1 & 3.4 & $0.5\pm 0.6$\\
                                  & 8\,TeV      & 120.0  & 3.3 & 1.0 & $1.9\pm 0.7$\\
                                  & 7 \& 8\,TeV & 125.0  & 2.8 & 2.3 & $1.3\pm0.5$& 124--233 & 137--261 \\
      \hline
      \multirow{4}{*}{Combined}   & 7\,TeV      & 126.5  & 3.6 & 3.2 & $1.2\pm 0.4$&         &   \\
                                  & 8\,TeV      & 126.5  & 4.9 & 3.8 & $1.5\pm 0.4$&         &   \\ 
                                  & \multirow{2}{*}{7 \& 8\,TeV} & \multirow{2}{*}{126.5}  & \multirow{2}{*}{6.0} &\multirow{2}{*}{4.9} & \multirow{2}{*}{$1.4\pm 0.3$} & \lowerExpNoGeV--\upperExpNoGeV & \lowerlowerObsNoGeV--\upperlowerObsNoGeV, \lowerObsNoGeV--\upperObsNoGeV \\
                                  &                              &                         &                      & &                               &                   113--532 (*) & 113--114, 117--121, 132--527 (*) \\
      \hline\hline
    \end{tabular}} 
  \end{table*}

\section{Correlated systematic uncertainties}
\label{sec:CombSyst}

The individual search channels that enter the combination are
summarised in Table~\ref{tab:channels}. 

The main uncorrelated systematic
uncertainties 
are described in
Sections~\ref{sec:h4l}--\ref{sec:hww} for the \hZZllll, \hgg\ and
\hWWlnln\ channels and in Ref.~\cite{paper2012prd} for the other
channels. They include the background normalisations or background
model parameters from control regions or sidebands, the Monte
Carlo simulation statistical uncertainties and the theoretical
uncertainties affecting the background processes.

The main sources of correlated systematic uncertainties are the
following.

1. {\it Integrated luminosity:} The uncertainty on the integrated
luminosity is considered as fully correlated among channels and
amounts to $\pm$3.9\%\ for the 7\,TeV data~\cite{Aad:2011dr,
LuminosityCONF}, except for the \hZZllll\ and \hgg\ channels which were
re-analysed; the uncertainty is $\pm$1.8\%\ ~\cite{ATLAS-CONF-2012-080} for these channels. 
The uncertainty is $\pm$3.6\%\ for the 8\,TeV data.

2. {\it Electron and photon trigger identification:} The uncertainties in the
trigger and identification efficiencies are treated as fully
correlated for electrons and photons.

3. {\it Electron and photon energy scales:} The electron and photon
energy scales in the \hZZllll\ and  \hgg\ channels are described by
five parameters, which provide a detailed account of the sources of
systematic uncertainty.  They are related to the calibration method,
the presampler energy scale in the barrel and end-cap
calorimeters, and the material description upstream of the calorimeters.

4. {\it Muon reconstruction:} The uncertainties affecting muons are separated into those related
to the ID and MS, in order to
obtain a better description of the correlated effects among channels
using different muon identification criteria and different ranges of
muon $p_{\rm T}$. 

5. {\it Jet energy scale and missing transverse energy:} The jet
energy scale and jet energy resolution are affected by
uncertainties which depend on the $p_{\rm T}$, $\eta$, and flavour of
the jet. 
A simplified scheme is used in which independent JES and JER nuisance
parameters are associated with final states with significantly different
kinematic selections and sensitivity to scattering processes with
different kinematic distributions or flavour composition.  This scheme
includes a specific treatment for $b$-jets. The sensitivity of the
results to various assumptions about the correlation between these
sources of uncertainty has been found to be negligible.
An uncorrelated component of the uncertainty on $E_{\rm T}^{\rm miss}$ is
included, in addition to the JES uncertainty, which is due to low energy jet activity not associated with reconstructed physics objects.

6. {\it Theory uncertainties:} Correlated theoretical uncertainties
affect mostly the signal predictions. The QCD scale uncertainties for
\mH=125~\GeV\ amount to $^{+7\%}_{-8\%}$ for the ggF process,
$\pm1$\%\ for the VBF and $WH/ZH$ processes, and
$^{+4\%}_{-9\%}$ for the $t\bar{t}H$ process~\cite{LHCHiggsCrossSectionWorkingGroup:2011ti,Dittmaier:2012vm};
the small dependence of these uncertainties on $m_H$ is taken into account.
The uncertainties on the predicted branching ratios amount to $\pm5\%$.
The uncertainties related to
the parton distribution functions amount to $\pm 8$\%\ for the
predominantly gluon-initiated ggF and $t\bar{t}H$ processes, and $\pm 4$\%\
for the predominantly quark-initiated VBF and $WH/ZH$
processes~\cite{Botje:2011sn,Lai:2010vv,Martin:2009iq,Ball:2011mu}.
The theoretical uncertainty associated
with the exclusive Higgs boson production process with additional jets
in the \hgg, \hWWlnln\ and \htt\ channels 
is estimated using the prescription of
Refs.~\cite{LHC-HCG,Stewart:2011cf,Dittmaier:2012vm}, with the
noticeable difference that an explicit calculation of the gluon-fusion
process at NLO using MCFM~\cite{Campbell:2006xx} in the 2-jet
category reduces the uncertainty on this non-negligible contribution to
25\,\%.
An additional theoretical uncertainty on the signal
normalisation of $\pm$150\%$\times$(\mH/TeV)$^3$
(e.g.\ $\pm 4\%$ for $\mH=300$~\GeV)
accounts for effects related to off-shell Higgs boson
production and interference with other SM processes~\cite{Dittmaier:2012vm}.

Sources of systematic uncertainty that affect both the 7\,TeV and the 8\,TeV data are taken
as fully correlated. The uncertainties on background estimates based on control samples in the data
are considered uncorrelated between the 7\,TeV and 8\,TeV data.


\begin{figure}[t!]
  \begin{center}
    \includegraphics[height=.5\textheight]{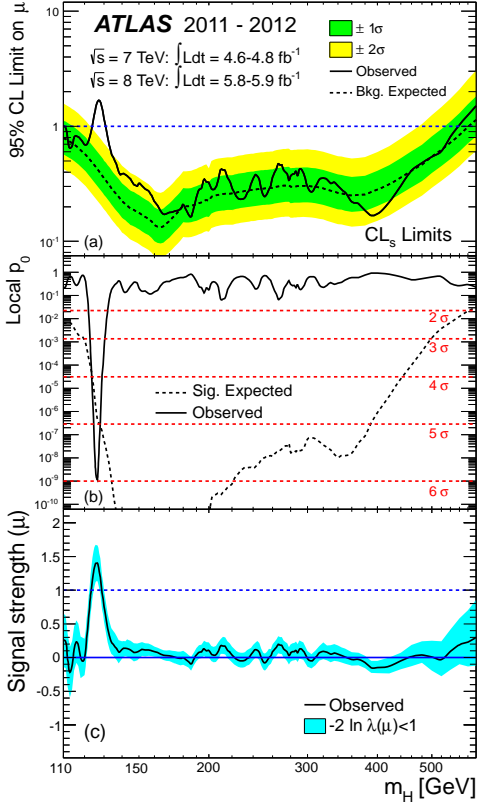}
\vspace*{-0.5cm}
\caption{Combined search results: (a) The observed (solid) 95\%~CL limits on the signal
  strength as a function of \mh\ and the expectation (dashed) under
  the background-only hypothesis. The dark and light shaded bands show the $\pm 1\sigma$
  and $\pm 2\sigma$ uncertainties on the background-only expectation. (b) The observed (solid) local $p_0$
  as a function of \mh\ and the expectation (dashed) for a SM Higgs boson
  signal hypothesis ($\mu=1$) at the given mass. (c) The best-fit signal strength $\hat\mu$ as
  a function of \mh. The band indicates the approximate 68\%\ CL interval around the fitted value.}
  \label{fig:CLsetal}
  \end{center}
\end{figure}

\section{Results\label{sec:Results}}

The addition of the 8\,TeV data for the \hZZllll, \hgg\ and $\hwwenmun$ channels, as well as the
improvements to the analyses of the 7\,TeV data in the first two of these channels, bring 
a significant gain in sensitivity in the low-mass region with respect to the
previous combined search~\cite{paper2012prd}.

\subsection{Excluded mass regions}

The combined 95\%~CL exclusion limits on the production of the SM Higgs
boson, expressed in terms of the signal strength parameter $\mu$, are
shown in Fig.~\ref{fig:CLsetal}(a) as a function of \mh.
The expected 95\%~CL exclusion region covers the \mH\ range from
\lowerExp\ to \upperExp. The observed 95\%~CL exclusion regions are
\lowerlowerObsNoGeV --\upperlowerObs\ and \lowerObsNoGeV --\upperObs.
Three mass regions are excluded at 99\%~CL,
113--114, 117--121 and 132--527\,GeV, while the
expected exclusion range at 99\%\,CL is 113--532\,GeV.

\begin{figure}[!t]
\begin{center}
\includegraphics[height=.5\textheight]{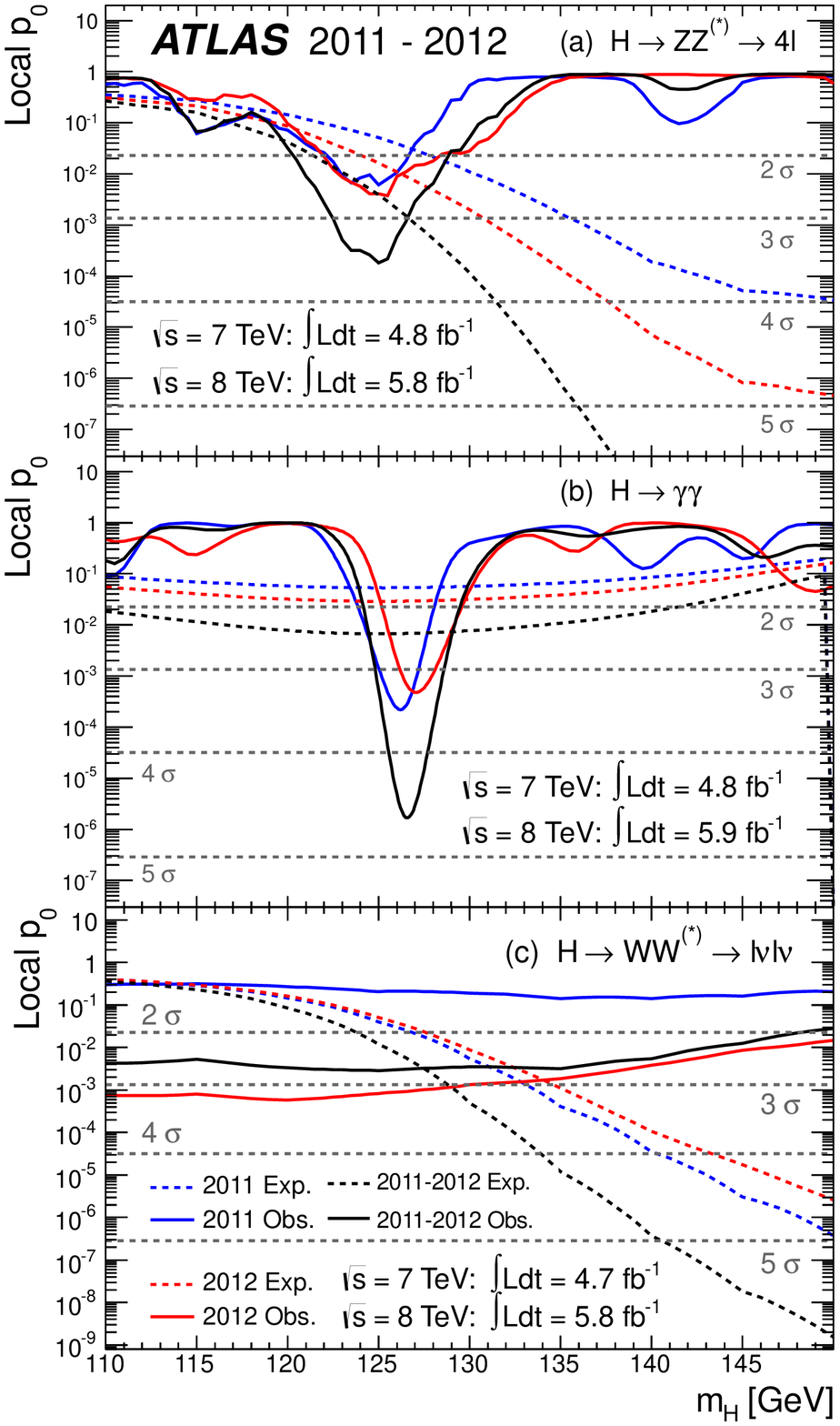}
\vspace*{-0.5cm}
\caption{The observed local $p_0$ as a
  function of the hypothesised Higgs boson mass for the
   (a) \htollll, (b) \hgg\ and (c) \hwwlnln\ channels. 
  The dashed curves show the expected local
  $p_0$ under the hypothesis of a SM Higgs boson signal at
  that mass.
  Results are shown separately for the $\sqrt{s}=7$\,TeV data
  (dark, blue), the $\sqrt{s}=8$\,TeV data (light, red), and their combination (black).}
\label{fig:allp0}
\end{center}
\end{figure}

\subsection{Observation of an excess of events}

An excess of events is observed near \mh$=$126\,GeV in the 
\hZZllll\ and \hgg\ channels, both of which provide fully reconstructed
candidates with high resolution in invariant mass, as shown in
Figures~\ref{fig:allp0}(a) and \ref{fig:allp0}(b). These excesses
are confirmed by the highly sensitive but low-resolution 
$\hwwlnln$ channel, as shown in Fig.~\ref{fig:allp0}(c).

\begin{figure}[!tb]
  \centering
  \includegraphics[width=\linewidth]{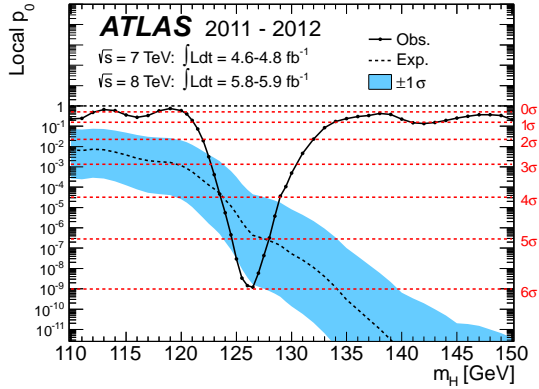}
\vspace*{-0.5cm}
\caption{The observed (solid) local $p_0$ as a
  function of \mh\ in the low mass range. The dashed curve shows the expected local
  $p_0$ under the hypothesis of a SM Higgs boson signal at
  that mass with its $\pm 1\sigma$ band. The horizontal dashed
  lines indicate the $p$-values corresponding to significances of 1 to
  6~$\sigma$.}\label{fig:CLb}
\end{figure}

The observed local $p_0$ values from the combination of channels, using the
asymptotic approximation, are shown as a function of \mh\ in
Fig.~\ref{fig:CLsetal}(b) for the full mass range and in Fig.~\ref{fig:CLb} for the low mass range.

The largest local significance for the combination of the 7 and 8~TeV data is found
for a SM Higgs boson mass hypothesis of \mH=126.5\,\GeV, where it reaches
\significance, with an expected value in the presence of a SM Higgs boson
signal at that mass of \expectedsignificance\ (see also Table~\ref{tab:statsummary}).  
For the 2012 data alone, the
maximum local significance for the \hZZllll, \hgg\ and $\hwwenmun$ channels
combined is 4.9\,$\sigma$, and occurs at $m_H=126.5$\,\GeV\ (3.8\,$\sigma$
expected).

The significance of the excess is mildly sensitive to uncertainties
in the energy resolutions and energy scale systematic uncertainties for photons and
electrons; the effect of the muon energy scale systematic
uncertainties is negligible.  The presence of these uncertainties,
evaluated as described in Ref.~\cite{ATLAS-CONF-2012-093}, reduces the local
significance to 5.9\,$\sigma$.

The global significance of a local \significanceESS\ excess anywhere
in the mass range 110--600\,GeV is estimated to be approximately
5.1\,$\sigma$, increasing to 5.3\,$\sigma$ in the range 110--150\,\GeV ,
which is approximately the mass range not excluded at the 99\%\ CL by
the LHC combined SM Higgs boson search~\cite{lhcCombination} and the
indirect constraints from the global
fit to precision electroweak measurements~\cite{lepew:2010vi}.

\subsection{Characterising the excess}

The mass of the observed new particle is estimated using the profile
likelihood ratio $\lambda(m_H)$ for \hZZllll\ and \hgg, the two
channels with the highest mass resolution.  The signal strength is
allowed to vary independently in the two channels, although the result
is essentially unchanged when restricted to the SM hypothesis $\mu=1$.
The leading sources of systematic uncertainty come from
the electron and photon energy scales and resolutions.
The resulting estimate for the mass of the observed particle is
$\massresultStatSys$.

\begin{figure}
\center
\includegraphics[width=.4\textwidth]{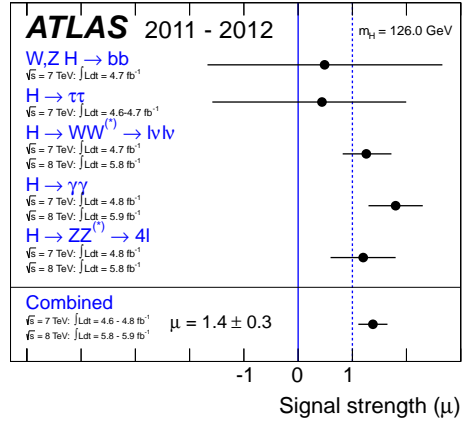}
\vspace*{-0.5cm}
\caption{Measurements of the signal strength parameter $\mu$ for \mh=126\,GeV\ for the individual
channels and their combination.}
\label{fig:mubarchart}
\end{figure}

The best-fit signal strength $\hat{\mu}$ is
shown in Fig.~\ref{fig:CLsetal}(c) as a function of \mh.  The observed excess corresponds
to $\hat{\mu} = 1.4\pm 0.3$ for $\mH=126$\,\GeV, which
is consistent with the SM Higgs boson hypothesis $\mu=1$. A summary of the individual and
combined best-fit values of the strength parameter for a SM Higgs
boson mass hypothesis of 126\,GeV is shown in Fig.~\ref{fig:mubarchart}, while
more information about the three main channels is provided in Table~\ref{tab:statsummary}.

In order to test which values of the strength and mass 
of a signal hypothesis are simultaneously consistent with the data, the
profile likelihood ratio $\lambda(\mu,m_H)$ is used.
In the presence of a strong signal, it will produce closed contours around the best-fit
point $(\hat\mu,\hat m_H)$, while in the absence of a signal the
contours will be upper limits on $\mu$ for all values of \mh.

\begin{figure}[t]
\center
\includegraphics[width=.48\textwidth]{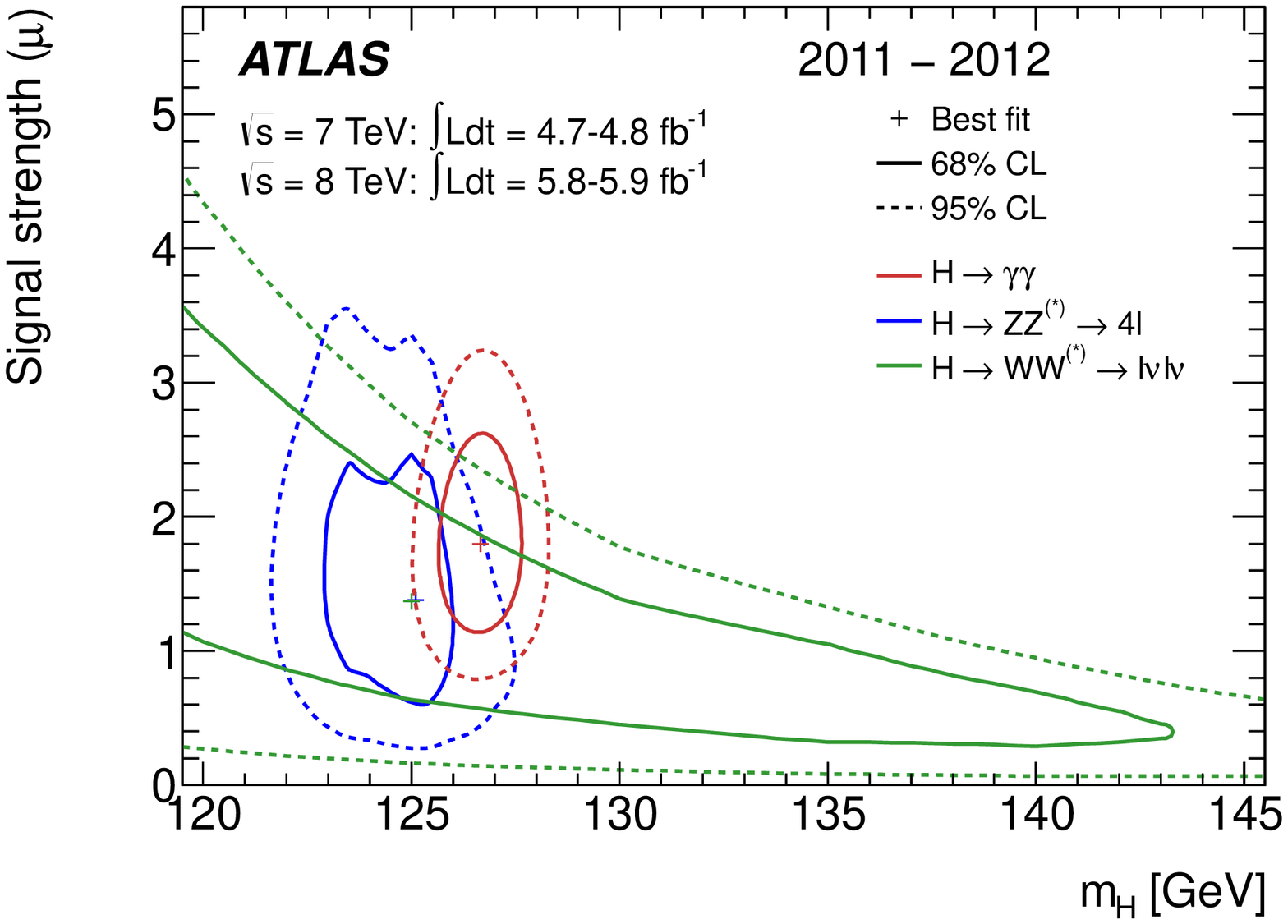}
\vspace*{-0.5cm}
\caption{Confidence intervals in the $(\mu,\mh)$ plane for the \hZZllll,
  \hgg, and \hWWlnln\ channels, including all systematic uncertainties. The markers indicate the maximum likelihood estimates $(\hat\mu,\hat{m}_H)$ in the corresponding channels (the maximum likelihood estimates for \hZZllll\
and \hWWlnln\ coincide).}
\label{fig:ggZZcontour}
\end{figure}

Asymptotically, the test statistic $-2\ln \lambda(\mu,m_H)$ is
distributed as a $\chi^2$ distribution with two degrees of freedom.
The resulting 68\% and 95\%
CL contours for the \hgg\ and \hWWlnln\ channels are shown in
Fig.~\ref{fig:ggZZcontour}, where the asymptotic approximations have
been validated with ensembles of pseudo-experiments. 
Similar 
contours
for the \hZZllll\ channel are also shown in
Fig.~\ref{fig:ggZZcontour}, although they are only approximate confidence intervals  
due to the smaller number of candidates in this channel.
These contours in the $(\mu,\mh)$ plane take
into account uncertainties in the energy scale and resolution.

The probability for a single Higgs boson-like particle
to produce resonant mass peaks in the \hZZllll\ and \hgg\ channels separated 
by more than the observed mass difference,
allowing the signal strengths to vary independently, is about 8\%.

The contributions from the different production modes in the \hgg\
channel have been studied in order to assess any tension between the
data and the ratios of the production cross sections predicted in the Standard Model.
A new signal strength parameter $\mu_i$ is introduced for each
production mode, defined by $\mu_i = \sigma_i / \sigma_{i, \rm {\rm SM}}$.
In order to determine the values of $(\mu_i,\mu_j)$ that are
simultaneously consistent with the data, the profile likelihood ratio
$\lambda(\mu_i,\mu_j)$ is used with the measured mass treated as a nuisance
parameter.

\begin{figure}[t]
 \centering
    \includegraphics[width=.48\textwidth]{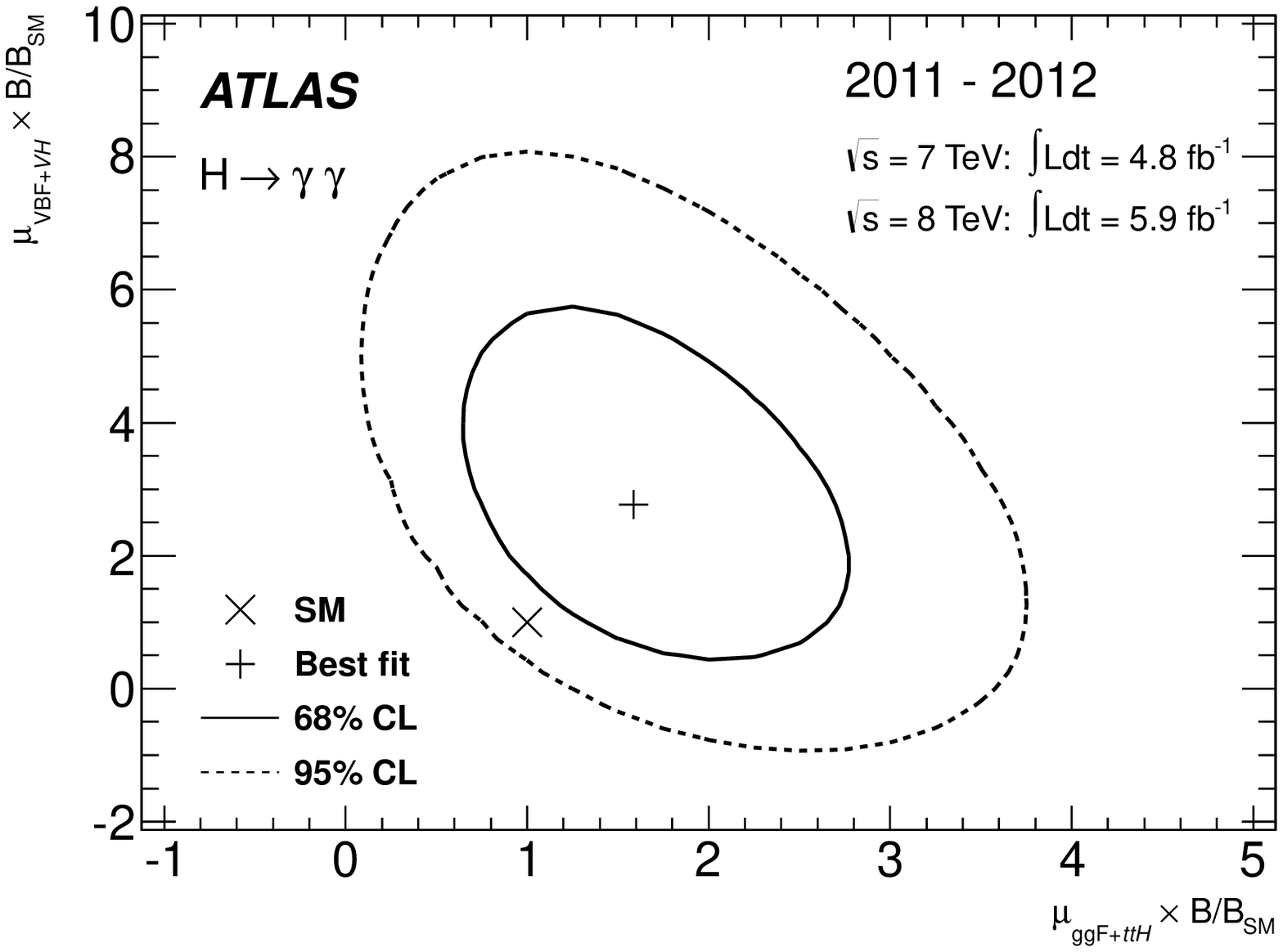}
    \vspace*{-0.5cm}
    \caption{Likelihood contours for the \hgg\
      channel in the $(\mu_{\mathrm{ggF}+t\bar{t}H}, \mu_{\mathrm{VBF}+VH})$ plane including the branching ratio factor $B/B_{\rm {\rm SM}}$.
      The quantity $\mu_{\mathrm{ggF}+t\bar{t}H}$ ($\mu_{\mathrm{VBF}+VH}$) is a common scale
      factor for the ggF and $t\bar{t}H$ (VBF and $VH$) production
      cross sections. The best fit to the data (+) and 68\%\ (full) and 95\%\ (dashed) CL
      contours are also indicated, as well as the SM expectation ($\times$).}
    \label{fig:gamgamProdContourProf}
 \end{figure}

Since there are four Higgs boson production modes at the LHC, two-dimensional
contours require either some $\mu_i$ to be fixed, or multiple
$\mu_i$ to be related in some way.  Here, $\mu_{\mathrm{ggF}}$ and $\mu_{t\bar{t}H}$
have been grouped together as they scale with the $t\bar{t}H$ coupling in
the SM, and are denoted by the common parameter
$\mu_{\mathrm{ggF}+t\bar{t}H}$. Similarly, $\mu_\mathrm{VBF}$ and $\mu_{VH}$ have been
grouped together as they scale with the $WWH/ZZH$ coupling in the SM,
and are denoted by the common parameter $\mu_{\mathrm{VBF}+VH}$. 
Since the distribution of signal events among the 10 categories
of the  \hgg\ search is sensitive to these factors, constraints 
in the plane of $\mu_{\mathrm{ggF}+t\bar{t}H} \times
B/B_{\rm {\rm SM}}$ and $\mu_{\mathrm{VBF}+VH} \times B/B_{\rm {\rm SM}}$, where $B$ is the branching ratio for \hgg, can be
obtained (Fig.~\ref{fig:gamgamProdContourProf}). Theoretical uncertainties are included so that the consistency with the SM expectation can be quantified.
The data are compatible with the SM expectation at the 1.5\,$\sigma$ level. 

\section{Conclusion\label{sec:Conclusion}}

Searches for the Standard Model Higgs boson have been performed in the
\htollll, \hgg\  and \hWWenmun\ channels with the ATLAS experiment at the
LHC using 5.8--5.9~\infb\ of $pp$ collision data recorded during April to
June 2012 at a centre-of-mass energy of 8\,TeV. These results are
combined with earlier results~\cite{paper2012prd}, which are
based on an integrated luminosity of 4.6--4.8~\infb\ recorded in 2011
at a centre-of-mass energy of 7\,TeV, except for the 
\htollll\ and \hgg\ channels, which have been updated with
the improved analyses presented here.

The Standard Model Higgs boson is excluded at 95\%\ CL
in the mass range \lowerlowerObsNoGeV --\upperObs, except for the
narrow region \upperlowerObsNoGeV --\lowerObs. In this region, an
excess of events with significance \significanceESS, corresponding
to $p_0=1.7\times 10^{-9}$, is observed. The excess is driven
by the two channels with the highest mass resolution, 
\htollll\ and \hgg, and the equally sensitive but low-resolution \hWWlnln\ channel.
Taking into account the entire mass range of the search,  110--600\,GeV,
the global significance of the excess is $5.1\,\sigma$, which corresponds
to $p_0=1.7\times 10^{-7}$.

These results provide conclusive evidence for the discovery of a new 
particle with mass $\massresultStatSys$. The signal strength parameter $\mu$
has the value $1.4\pm 0.3$ at the fitted mass,
which is consistent with the SM Higgs boson hypothesis $\mu=1$.
The decays to pairs of vector 
bosons whose net electric charge is zero identify the new particle as a neutral 
boson.
The observation in the  diphoton channel disfavours the spin-1 
hypothesis~\cite{landau1948,yang1950}. 
Although these  results are compatible with the hypothesis that the new particle is the 
Standard Model Higgs boson, more data are needed to assess its nature in detail.


\section*{Acknowledgements}

The results reported in this Letter would not have been possible without the
outstanding performance of the LHC. We warmly thank CERN and the entire LHC exploitation team, including 
the operation, technical and infrastructure groups, and all the people who have contributed to 
the conception, design and construction of this superb accelerator.
We thank also the support staff at our institutions without whose excellent contributions ATLAS could
not have been successfully constructed or operated so efficiently. 

We acknowledge the support of ANPCyT, Argentina; YerPhI, Armenia; ARC,
Australia; BMWF, Austria; ANAS, Azerbaijan; SSTC, Belarus; CNPq and FAPESP,
Brazil; NSERC, NRC and CFI, Canada; CERN; CONICYT, Chile; CAS, MOST and NSFC,
China; COLCIENCIAS, Colombia; MSMT CR, MPO CR and VSC CR, Czech Republic;
DNRF, DNSRC and Lundbeck Foundation, Denmark; EPLANET and ERC, European Union;
IN2P3-CNRS, CEA-DSM/IRFU, France; GNAS, Georgia; BMBF, DFG, HGF, MPG and AvH
Foundation, Germany; GSRT, Greece; ISF, MINERVA, GIF, DIP and Benoziyo Center,
Israel; INFN, Italy; MEXT and JSPS, Japan; CNRST, Morocco; FOM and NWO,
Netherlands; RCN, Norway; MNiSW, Poland; GRICES and FCT, Portugal; MERYS
(MECTS), Romania; MES of Russia and ROSATOM, Russian Federation; JINR; MSTD,
Serbia; MSSR, Slovakia; ARRS and MVZT, Slovenia; DST/NRF, South Africa;
MICINN, Spain; SRC and Wallenberg Foundation, Sweden; SER, SNSF and Cantons of
Bern and Geneva, Switzerland; NSC, Taiwan; TAEK, Turkey; STFC, the Royal
Society and Leverhulme Trust, United Kingdom; DOE and NSF, United States of
America.

The crucial computing support from all WLCG partners is acknowledged
gratefully, in particular from CERN and the ATLAS Tier-1 facilities at
TRIUMF (Canada), NDGF (Denmark, Norway, Sweden), CC-IN2P3 (France),
KIT/GridKA (Germany), INFN-CNAF (Italy), NL-T1 (Netherlands), PIC (Spain),
ASGC (Taiwan), RAL (UK) and BNL (USA) and in the Tier-2 facilities
worldwide.

\bibliographystyle{atlasBibStyleWithTitle} 
\providecommand{\href}[2]{#2}\begingroup\raggedright\endgroup

\clearpage
\onecolumn
\input{atlas_authlist}
 
\end{document}

%% file: atlas_authlist.tex
\begin{flushleft}
{\Large The ATLAS Collaboration}

\bigskip

G.~Aad$^{\rm 48}$,
T.~Abajyan$^{\rm 21}$,
B.~Abbott$^{\rm 111}$,
J.~Abdallah$^{\rm 12}$,
S.~Abdel~Khalek$^{\rm 115}$,
A.A.~Abdelalim$^{\rm 49}$,
O.~Abdinov$^{\rm 11}$,
R.~Aben$^{\rm 105}$,
B.~Abi$^{\rm 112}$,
M.~Abolins$^{\rm 88}$
% ...
\bigskip

% ...
$^{11}$ Institute of Physics, Azerbaijan Academy of Sciences, Baku, Azerbaijan\\
$^{12}$ Institut de F\'{i}sica d'Altes Energies and Departament de F\'{i}sica de la Universitat Aut\`{o}noma de Barcelona and ICREA, Barcelona, Spain\\
% ...
$^{21}$ Physikalisches Institut, University of Bonn, Bonn, Germany\\
% ...
$^{48}$ Fakult\"{a}t f\"{u}r Mathematik und Physik, Albert-Ludwigs-Universit\"{a}t, Freiburg, Germany\\
$^{49}$ Section de Physique, Universit\'{e} de Gen\`{e}ve, Geneva, Switzerland\\
% ...
$^{88}$ Department of Physics and Astronomy, Michigan State University, East Lansing MI, United States of America\\
% ...
$^{105}$ Nikhef National Institute for Subatomic Physics and University of Amsterdam, Amsterdam, Netherlands\\
% ...
$^{111}$ Homer L. Dodge Department of Physics and Astronomy, University of Oklahoma, Norman OK, United States of America\\
$^{112}$ Department of Physics, Oklahoma State University, Stillwater OK, United States of America\\
% ...
$^{115}$ LAL, Universit\'{e} Paris-Sud and CNRS/IN2P3, Orsay, France\\
% ...
\end{flushleft}
